# Large Language Models as Subpopulation Representative Models: A Review

Gabriel Simmons
University of California, Davis
Department of Computer Science
gsimmons@ucdavis.edu

Christopher Hare
University of California, Davis
Department of Political Science
cdhare@ucdavis.edu

## Abstract

Of the many commercial and scientific opportunities provided by large language models (LLMs; including Open AI's ChatGPT, Meta's LLaMA, and Anthropic's Claude), one of the more intriguing applications has been the simulation of human behavior and opinion. LLMs have been used to generate human simulcra to serve as experimental participants, survey respondents, or other agents—with outcomes that often parallel the observed behavior of their genuine human counterparts. Here, we consider the feasibility of using LLMs to estimate subpopulation representative models (SRMs). SRMs could provide an alternate or complementary way to measure public opinion among demographic, geographic, or political segments of the population. However, the introduction of new technology to the socio-technical infrastructure does not come without risk. We provide an overview of behavior elicitation techniques for LLMs, and a survey of existing SRM implementations. We offer frameworks for the analysis, development, and practical implementation of LLMs as SRMs, consider potential risks, and suggest directions for future work.

## 1. Introduction

In 1961, V.O. Key wrote in *Public Opinion and American Democracy* that "to speak with precision of public opinion is a task not unlike coming to grips with the Holy Ghost."[1]. Even at this early stage in modern public opinion research, social scientists were already applying computing power to unravel the ethereal qualities of individuals' attitudes. That same year, Thomas B. Morgan penned an article in *Harper's* entitled "The People Machine".[2] This article detailed the efforts of a "secret weapon" used by the Kennedy campaign in the 1960 presidential election: a "People Machine" that harnessed computing power and reams of public opinion polling data to divide the American electorate into 480 voter types (e.g., "Midwestern, rural, Protestant, lower income, female").[3] The so-called People Machine (which was the flagship product of the newfound Simulmatics Corporation) pooled survey data to analyze each group's policy attitudes and simulate election outcomes under various hypothetical situations. For instance, how would public opinion shift if Kennedy were to take a stronger position in support of civil rights, or to place greater emphasis on his Catholicism and discrimination against religious minorities?

Morgan's article is a fascinating time capsule, fascinated but also fearful about technological change and its possibility to remake liberal democracy in the postwar era. Indeed, the piece includes a quote from Harold Laswell referring to the effort as "the A-bomb of the social sciences." While it is obvious in hindsight that the technology of the period was woefully underpowered to simulate the complexities of voter behavior, the concerns expressed in "The People Machine" are enduring and worth revisiting when considering how new technologies

---

[1] V.O. Key Jr., *Public Opinion and American Democracy* (New York: Knopf, 1961), p. 8.

[2] Thomas B. Morgan, "The People-Machine," *Harper's* 222 (1961): 53–57.

[3] For a comprehensive history of the Simulmatics Corporation and its intersection with social science, we highly recommend LePore, *If Then: How the Simulmatics Corpora-* *tion Invented the Future* and Pool, Abelson, and Popkin, *Candidates, Issues, and Strategies: A Computer Simulation of the 1960 and 1964 Presidential Elections.*

might be used to measure, study, and persuade public opinion.

The modern era has witnessed a proliferation of innovative techniques for aggregating community sentiment, reflecting the growing complexity of social structures, and the demand for more efficient and representative means of communication. The emergence of mass media, opinion polling, television, and the internet have all revolutionized the ways people express their views and engage with the opinions of others.[4]

At the same time, many of the challenges confronting public opinion research persist. For instance, response rates to telephone surveys have collapsed over recent decades, and now routinely fall below 10% in "gold standard" surveys fielded by organizations like the Pew Research Center and the American National Election Studies.[5] Not only does this make it more difficult to collect survey data with adequate sample sizes for subgroup analysis, but it also exacerbates nonresponse bias when response rates vary in nonrandom ways across subpopulations. Nonresponse bias is especially pernicious when it is related to relevant yet unobservable characteristics (e.g., propensity to vote for Republican candidates), but remains problematic even when deviations in response rates occur along observable characteristics (e.g., race or age). For example, an influential thread of work in survey methodology is generally pessimistic about the ability of various weighting schemes to correct for known sampling biases when using nonprobability samples, as is the case for virtually every online survey panel.[6] While some of these concerns may be exaggerated[7], mainstream survey research nonetheless continues to struggle in collecting adequately large and representative samples to gauge public opinion at subnational and subgroup levels. For instance, polling errors in the 2016 presidential election were most pronounced in a handful of Midwestern Rust Belt states and among college-educated white voters.[8]

Certainly, the fundamental difficulties associated with measuring subpopulation sentiment are well understood and documented. In addition to concerns regarding sample representativeness, even surveys with superficially large numbers of respondents encounter the so-called sparse cell problem. Namely, observations quickly grow sparse when we repeatedly stratify the sample along discrete characteristics (incurring the "curse of dimensionality" especially familiar to those in the machine and statistical learning communities). Consequently, conducting valid inference on subpopulations becomes increasingly difficult. At present, social scientists most frequently address this problem with methods that exploit aggregate patterns in the data to perform shrinkage on estimates for sparsely populated subgroups.[9] Specifically, multilevel regression and poststratification (MRP) and its variants are the state-of-the-art-method for estimating sub-and Empirical Evidence on Probability and Nonprobability Sample Survey Research," *Journal of Survey Statistics and Methodology* 8, no. 1 (2020): 4–36.

[4]For more comprehensive overviews of public opinion survey research and its evolution, see Herbst (1993) , Groves (2011) , Hillygus (2020) , and Berinsky (2017) .

[5]Thomas J Leeper, "Where Have the Respondents Gone? Perhaps We Ate Them All," *Public Opinion Quarterly* 83, S1 (2019): 280–288.

[6]Mario Callegaro et al., "A Critical Review of Studies Investigating the Quality of Data Obtained with Online Panels Based on Probability and Nonprobability Samples," in *Online Panel Research: A Data Quality Perspective*, eds. Mario Callegaro et al., New York: Wiley, 2014, 23–53., Carina Cornesse et al., "A Review of Conceptual Approaches

[7]Richard Hendra and Aaron Hill, "Rethinking Response Rates: New Evidence of Little Relationship between Survey Response Rates and Nonresponse Bias," *Evaluation Review* 43, no. 5 (2019): 307–330.

[8]Courtney Kennedy et al., "An Evaluation of the 2016 Election Polls in the United States," *Public Opinion Quarterly* 82, no. 1 (2018): 1–33.

[9]David K. Park, Andrew Gelman, and Joseph Bafumi, "Bayesian Multilevel Estimation with Poststratification: State-Level Estimates from National Polls," *Political Analysis* 12, no. 4 (2004): 375–385., Drew A. Linzer, "Reliable Inference in Highly Stratified Contingency Tables: Using Latent Class Models as Density Estimators," *Political Analysis* 19, no. 2 (2011): 173–187., Daniël W. van der Palm, L. Andries van der Ark, and Jeroen K. Vermunt, "Divisive Latent Class Modeling as a Density Estimation Method for Categorical Data," *Journal of Classification* 33, no. 1 (2016): 52–72.

group opinion with hierarchical/multilevel data, with MRP being an especially popular means of estimating subnational opinion (i.e., small-area estimation). Closely related methods such as group-level item response theory models also seek to leverage hierarchical structure in the data to exchange information or "borrow strength" across observations when estimating opinion for demographic and/or geographic constituencies.[10]

With these obstacles in mind, today's vast repository of user-generated content available online presents unparalleled opportunities for real-time acquisition and examination of public opinion, motivating the exploration of cutting-edge computational techniques to overcome the constraints of traditional polling and survey methods. In particular, recent work suggests that Large Language Models (LLMs) may be able to approximate the sentiment and opinion of human subpopulations or their individual members in useful ways.[11] However, this capability is still far from perfect.[12] This may not be surprising, since LLMs are trained on a vast amount of text data, typically collected from public websites on the Internet, including discussion forums and other places where community sentiment may be expressed. It stands to reason that those tasked with aggregating community sentiment would look to LLMs as a promising new means for this task. Initial work in this direction has identified several unique benefits in using LLMs, including more interactive and open-ended analyses[13], the potential for forecasting[14], decreased cost to conduct social research[15], and the potential to bridge disagreements.[16] While LLMs are not necessarily a panacea for problems like data sparsity, they offer a promising way to leverage richer and more expressive language data to recover more valid and reliable estimates of subpopulation opinion.

This review draws together a body of literature that uses LLMs to approximate subpopulation characteristics, or aggregate subpopulation sentiment, under the umbrella of Subpopulation Representative Models (SRMs). In Section 2, we provide background on LLMs and the techniques used to steer their behavior. A definition for Subpopulation Representative Models is then provided in Section 3, and a description of the development life cycle for SRMs in Section 4. In Section 5, the existing implementations of Subpopulation Representative Models from referred academic publications, non-refereed academic publications, and other sources are reviewed, summarizing key implementation details. A discussion and recommendations for future work are then provided in Section 6.

[10] See, e.g., Caughey and Warshaw (2015) .

[11] Gati Aher, Rosa I. Arriaga, and Adam Tauman Kalai, *Using Large Language Models to Simulate Multiple Humans and Replicate Human Subject Studies*, arXiv, 2023, https://doi.org/10.48550/arXiv.2208.10264., Lisa P. Argyle et al., "Out of One, Many: Using Language Models to Simulate Human Samples," *Political Analysis* 31, no. 3 (2023): 337–351., Lewis D. Griffin et al., *Susceptibility to Influence of Large Language Models*, arXiv, 2023, https://doi.org/10.48550/arXiv.2303.06074., Hang Jiang et al., "CommunityLM: Probing Partisan Worldviews from Language Models," in *Proceedings of the 29th International Conference On Computational Linguistics* (Gyeongju, Republic of Korea: International Committee on Computational Linguistics, 2022), 6818–6826, https://aclanthology.org/2022.coling-1.593.

[12] Shibani Santurkar et al., *Whose Opinions Do Language Models Reflect?*, arXiv, 2023, https://doi.org/10.48550/arXiv.2303.17548.

[13] Philip G. Feldman, Shimei Pan, and James Foulds, "The Keyword Explorer Suite: A Toolkit for Understanding Online Populations," in *Companion Proceedings of the 28th International Conference on Intelligent User Interfaces* (New York: Association for Computing Machinery, 2023), 21–24, https://doi.org/10.1145/3581754.3584122.

[14] Eric Chu et al., *Language Models Trained on Media Diets Can Predict Public Opinion*, arXiv, 2023, https://doi.org/10.48550/arXiv.2303.16779.

[15] Shriphani Palakodety, Ashiqur R. KhudaBukhsh, and J. Carbonell, *Mining Insights from Large-Scale Corpora Using Fine-Tuned Language Models*, Amsterdam: IOS Press, 2020, 1890–1897., Argyle et al., "Out of One, Many: Using Language Models to Simulate Human Samples," 337–351.

[16] Michiel A. Bakker et al., "Fine-Tuning Language Models to Find Agreement among Humans with Diverse Preferences," *Advances in Neural Information Processing Systems* 35 (2022): 38176–38189.

## 2. Background on Large Language Models

This section provides a comprehensive overview of Large Language Models, their architecture, training processes, and methods for eliciting desired outputs. We include this section as a resource for practitioners who use LLMs as SRMs but may not have familiarity with the underlying technology or its history in the context of machine learning.

LLMs attempt to model a conditional probability distribution over tokens (words or sub-word pieces) from a fixed vocabulary. In an LLM, this distribution is parameterized by a neural network, usually with billions of trainable parameters. LLMs have also been referred to as foundation models, although this category also includes vision and multimodal models.[17]

One of the key ideas that has emerged from the study of large neural networks is *transfer learning*: the idea that models trained on one task can be effectively used for related tasks with inexpensive adaptation. Transfer learning most commonly follows a pattern where models first undergo an initial training phase based on data that is not targeted towards any particular task or domain. This is often referred to as *pre-training*, and frequently represents the bulk of the computational expense in developing the system. Next, a pre-trained model is adapted to a specific task or domain through an additional training phase, often referred to as *fine-tuning*. The computational and data requirements for fine-tuning are often orders of magnitude smaller than the requirements for pre-training. Additionally, given that language models operate on text sequences, the user can exert additional control over the model by providing text to condition the model towards the desired behavior, a technique known as *prompting*. Each of these techniques is described in more detail in Sections 2.1, 2.2, and 2.3, respectively.

LLMs are most often based on the Transformer architecture.[18] LLMs consist of an embedding layer, which maps each token in the vocabulary to a fixed-length vector known as an *embedding*, subsequent layers that manipulate the embeddings to form an updated numerical representation of the sequence, and an output layer that transforms the updated representation into a format useful for a particular task. For example, to use a language model for text generation, the output layer maps from the internal representation to a distribution over tokens in the vocabulary. Text is generated by sampling from this distribution, appending the sampled token to the input sequence, and continuing this process until a stop criterion is met.

LLMs inherit from a number of preceding approaches, notably recurrent neural networks, especially those employing attention[19] and word embedding models. Word embedding approaches such as word2vec[20] and GloVe[21] achieved vector representations for each discrete token in a fixed vocabulary. These embeddings are still often used to initialize the embedding layer in an LLM.[22] One of the key advancements made by the language modeling paradigm is to learn not only global word representations, but also contextualization (how to update the meanings of words based on their context) during pre-training. Additionally, LLMs vary from earlier approaches in sheer size. While a 300-dimensional `word2vec` model over a vocabulary with 50,000 tokens contains 15m parameters, typical LLMs contain over one billion learned parameters.

[17]Rishi Bommasani et al., *On the Opportunities and Risks of Foundation Models*, arXiv, 2022, https://doi.org/10.48550/arXiv.2108.07258.

[18]Ashish Vaswani et al., *Attention Is All You Need*, arXiv, 2017, https://doi.org/10.48550/arXiv.1706.03762.

[19]Dzmitry Bahdanau, Kyunghyun Cho, and Yoshua Bengio, *Neural Machine Translation by Jointly Learning to Align and Translate*, arXiv, 2016, https://doi.org/10.48550/arXiv.1409.0473.

[20]Tomas Mikolov et al., *Efficient Estimation of Word Representations in Vector Space*, arXiv, 2013, https://doi.org/10.48550/arXiv.1301.3781.

[21]Jeffrey Pennington, Richard Socher, and Christopher Manning, "GloVe: Global Vectors for Word Representation," in *Proceedings of the 2014 Conference on Empirical Methods in Natural Language Processing (EMNLP)* (Doha, Qatar: Association for Computational Linguistics, 2014), 1532–1543, https://doi.org/10.3115/v1/D14-1162.

[22]Jacob Devlin et al., "BERT: Pre-Training of Deep Bidirectional Transformers for Language Understanding," in *Proceedings of the 2019 Conference of the North American Chapter of the Association for Computational Linguistics: Human Language Technologies* (Minneapolis, MN: Association for Computational Linguistics, 2019), 4171–4186, https://doi.org/10.18653/v1/N19-1423.

The following sections provide an overview of methods for behavior elicitation from LLMs. We refer the reader to P. Liu et al. (2023) for more in-depth coverage.

## 2.1. Pre-training

LLMs begin with their parameters randomly initialized. The first step in the training process, *pre-training*, uses stochastic gradient descent to optimize the model parameters against a self-supervised learning objective on a large corpus of text data. This text is separated into sub-word pieces called *tokens*. When a model receives an input sentence as a string, this string is split into tokens, and each token is mapped to a fixed-length vector of continuous values, referred to as an *embedding*.

*Masked* language models, such as BERT[23], are trained to predict tokens in the text that have been hidden (masked) in naturally-occurring sequences of tokens. *Autoregressive* models like GPT-3[24] are trained to predict from some naturally occurring sequence of tokens some following sequence.

## 2.2. Fine-tuning

After pre-training, LLMs commonly undergo fine-tuning processes to improve their performance for specific applications, which results in further updates to the model parameters. There are several common practices for fine-tuning:

- **Self-supervised fine-tuning**: This method further trains the model on unlabeled domain-specific corpora, using the same objective used for pre-training.

- **Supervised fine-tuning**: In this approach, the model is tuned on a dataset of examples with ground-truth labels. Supervised fine-tuning tasks range from highly specific (e.g., sentiment analysis) to general (e.g., following instructions). *Instruction tuning* is a variant of supervised fine-tuning in which models are trained on examples that include task input, task output, and task descriptions in natural language.[25] Recent work demonstrates that fine-tuning on supervised data obtained from the model itself[26] or from another model may be useful for certain tasks.

- **Reward model fine-tuning**: This process involves training the language model based on scores from a learned reward model that evaluates the outputs of the language model. *Reinforcement Learning from Human Feedback* (RLHF) is an example of reward model training where human evaluations of model outputs are used to train the reward model.[27] An emerging line of work suggests that reward model fine-tuning may have fundamentally different training properties than supervised fine-tuning; reward model tuning may have a stronger preference to learn compositional structure in the underlying data.[28]

- **Efficient fine-tuning and adapters**: Due to the size and associated computational expense of fine-tuning LLMs, a variety of techniques have been developed to fine-tune LLMs at reduced cost. These techniques are broadly referred to as *Parameter-Efficient Fine-Tuning* (PEFT) methods. These approaches can be subdivided based on whether they update the model by changing its existing parameters or introducing new parameters. Sparse fine-tuning techniques typically select a subset of model parameters to adjust using a variety of selection heuristics.[29] In contrast, adapter methods leave the base model weights unchanged, adding a rel-

[23]Jacob Devlin et al., "BERT: Pre-Training of Deep Bidirectional Transformers for Language Understanding," in *Proceedings of the 2019 Conference of the North American Chapter of the Association for Computational Linguistics: Human Language Technologies* (Minneapolis, MN: Association for Computational Linguistics, 2019), 4171–4186, https://doi.org/10.18653/v1/N19-1423.

[24]Tom Brown et al., "Language Models Are Few-Shot Learners," *Advances in Neural Information Processing Systems* 33 (2020): 1877–1901.

atively small number of new parameters that are trained during fine-tuning.[30]

## 2.3. Inference-time techniques

After pre-training and fine-tuning, there are several additional techniques for eliciting the desired output from language models that may be applied at inference time:

- **Decoding strategies**: Given a sequence of tokens, an LLM can be used to generate text by selecting a token to follow the input sequence from the conditional probability distribution modeled by the LLM. This process is often referred to as *decoding*, and there are a number of decoding strategies commonly used in practice. The simplest (*greedy decoding*) selects the most likely next token at each step. *Sampling* refers to randomly selecting a token with the likelihood of token selection weighted by its output probability. Several methods are used to control sampling, two of the most popular are *top*-k sampling and *top*-p sampling (also known as nucleus sampling). Top-$k$ sampling selects only from the $k$ tokens with the highest probability from the output probability distribution, with $k$ being an integer value specified by the user. Top-$p$ sampling chooses the smallest set of tokens whose cumulative probability exceeds $p$, a continuous value between 0 and 1, also specified by the user. Another popular strategy is *beam search decoding*, where a number of possible continuations are maintained, and the continuation with the highest likelihood is selected after all continuations reach a stop condition. All of these techniques may be used in combination with a *temperature* parameter, which controls the randomness of the output probability distribution; higher temperature values lead to a flatter output probability distribution, while lower temperature values lead to an output distribution with more sharply defined modes. Constraints may also be placed on specific tokens or sequences. Constrained decoding refers to a technique in which outputs from the model must belong to a predefined set of acceptable output sequences. Logit biases refer to user-specified biases on specific tokens in the model vocabulary; positive biases make the model more likely to generate the biased tokens, and negative biases have the opposite effect.

- **Prompting**: This refers to the use of carefully designed input sequences that condition the model to produce the desired output.[31] These input sequences may be obtained from human involvement (through *prompt engineering*) or from an optimization process (through *prompt tuning*).[32] The "prompt" used to condition the model may be one or more discrete tokens (a *hard prompt*) or a hidden model representation (a *soft prompt*).[33] Prompts expressed as discrete tokens often include instructions or a roleplaying description. Few-shot prompting refers to a technique in which solved task examples are provided in the input.[34]

- **Iterative refinement**: *Chain-of-thought prompting* is a specific type of prompting that encourages the model to subdivide complex tasks into smaller steps, generate intermediate outputs, and use the results of this intermediate processing to solve the final task.[35] This can enable more accurate and coherent responses to complex queries or tasks.[36] Chain-of-thought prompting was an early example of a broader trend towards *iterative refinement*, where models are used to self-critique and iteratively update their own output.[37] These refined outputs can be used for further model fine-tuning.[38]

[25]Jason Wei et al., *Finetuned Language Models Are Zero-Shot Learners*, arXiv, 2022, https://doi.org/10.48550/arXiv.2109.01652.

[26]Yuntao Bai et al., *Constitutional AI: Harmlessness from AI Feedback*, arXiv, 2022, https://doi.org/10.48550/arXiv.2212.08073.

[27]Yuntao Bai et al., *Training a Helpful and Harmless Assistant with Reinforcement Learning from Human Feedback*, arXiv, 2022, https://doi.org/10.48550/arXiv.2204.05862.

[28]Emily Cheng, Mathieu Rita, and Thierry Poibeau, *On the Correspondence between Compositionality and Imitation in Emergent Neural Communication*, arXiv, 2023, https://doi.org/10.48550/arXiv.2305.12941.

[29]Zihao Fu et al., *On the Effectiveness of Parameter-Efficient Fine-Tuning*, arXiv, 2022, https://doi.org/10.48550/arXiv.2211.15583.

[30]Edward J. Hu et al., "LoRA: Low-Rank Adaptation of Large Language Models," in *International Conference on Learning Representations* (2022), https://openreview.net/forum?id=nZeVKeeFYf9.

- **Tool and knowledge augmentation**: Augmentations allow the model to access external sources of information or interact with external tools. *Retrieval augmentation* involves incorporating a retrieval mechanism that can access and integrate external knowledge from large-scale databases or documents, helping the model to provide more informed and contextually relevant responses.[39] ReAct[40] and Toolformer[41] are examples of augmentation techniques that allow LLMs to perform more specialized tasks using tools like calculators or programming language interpreters.

- **Agents**: LLM agents synthesize action plans based on their environment descriptions. Agents consist of an LLM augmented with interfaces to knowledge sources or tools, as well as memory modules that store compressed versions of the agent's history. Agents run in a loop, performing actions towards some goal, evaluating outputs from knowledge sources or tools, and taking further actions based on this new information.[42]

## 3. Subpopulation Representative Models

In this section, we define the concepts of subpopulation representative models and subpopulation representative behavior and introduce the criteria used to evaluate their estimation by large language models (LLMs).

- **Subpopulation representative models (SRMs)**: We define subpopulation representative models (SRMs) as "models that approximate to some useful degree certain characteristics of human subpopulations". We restrict the scope of this review to Subpopulation Representative Models implemented using LLMs. The distinction between the concept of subpopulation representative models and the closely related field of machine psychology is clarified in Section 4.7.

- **Subpopulation representative behavior (SRB)**: Subpopulation representative behavior (SRB) is "the approximation of subpopulation characteristics" - the intended behavior of a subpopulation representative model.

Our definitions of subpopulation representative modeling and behavior are informed by concepts introduced in the reviewed works. Argyle et al. (2023) proposed the concept of *algorithmic fidelity*,

[31]Pengfei Liu et al., "Pre-Train, Prompt, and Predict: A Systematic Survey of Prompting Methods in Natural Language Processing," *ACM Computing Surveys* 55, no. 9 (2023): 1–35.

[32]Yuxian Gu et al., "PPT: Pre-Trained Prompt Tuning for Few-Shot Learning," in *Proceedings of the 60th Annual Meeting Of the Association For Computational Linguistics (Volume 1: Long Papers)* (Dublin, Ireland: Association for Computational Linguistics, 2022), 8410–8423, https://aclanthology.org/2022.acl-long.576., Brian Lester, Rami Al-Rfou, and Noah Constant, *The Power of Scale for Parameter-Efficient Prompt Tuning*, arXiv, 2021, https://doi.org/10.48550/arXiv.2104.08691.

[33]Lester, Al-Rfou, and Constant.

[34]Brown et al., "Language Models Are Few-Shot Learners," 1877–1901.

[35]Jason Wei et al., "Chain-of-Thought Prompting Elicits Reasoning in Large Language Models," *Advances in Neural Information Processing Systems* 35 (2022): 24824–24837.

[36]Similar prompting strategies include Tree of Thoughts (ToT) (Yao, D. Yu, et al. 2023).

[37]William Saunders et al., *Self-Critiquing Models for Assisting Human Evaluators*, arXiv, 2022, https://doi.org/

10.48550/arXiv.2206.05802., Aman Madaan et al., *Self-Refine: Iterative Refinement with Self-Feedback*, arXiv, 2023, https://doi.org/10.48550/arXiv.2303.17651.

[38]Bai et al., *Constitutional AI: Harmlessness from AI Feedback*.

[39]Patrick Lewis et al., *Retrieval-Augmented Generation for Knowledge-Intensive NLP Tasks*, arXiv, 2021, https://doi.org/10.48550/arXiv.2005.11401.

[40]Shunyu Yao et al., *ReAct: Synergizing Reasoning and Acting in Language Models*, arXiv, 2023, https://doi.org/10.48550/arXiv.2210.03629.

[41]Timo Schick et al., *Toolformer: Language Models Can Teach Themselves to Use Tools*, arXiv, 2023, https://doi.org/10.48550/arXiv.2302.04761.

[42]Joon Sung Park et al., *Generative Agents: Interactive Simulacra of Human Behavior*, arXiv, 2023, https://doi.org/10.48550/arXiv.2304.03442.

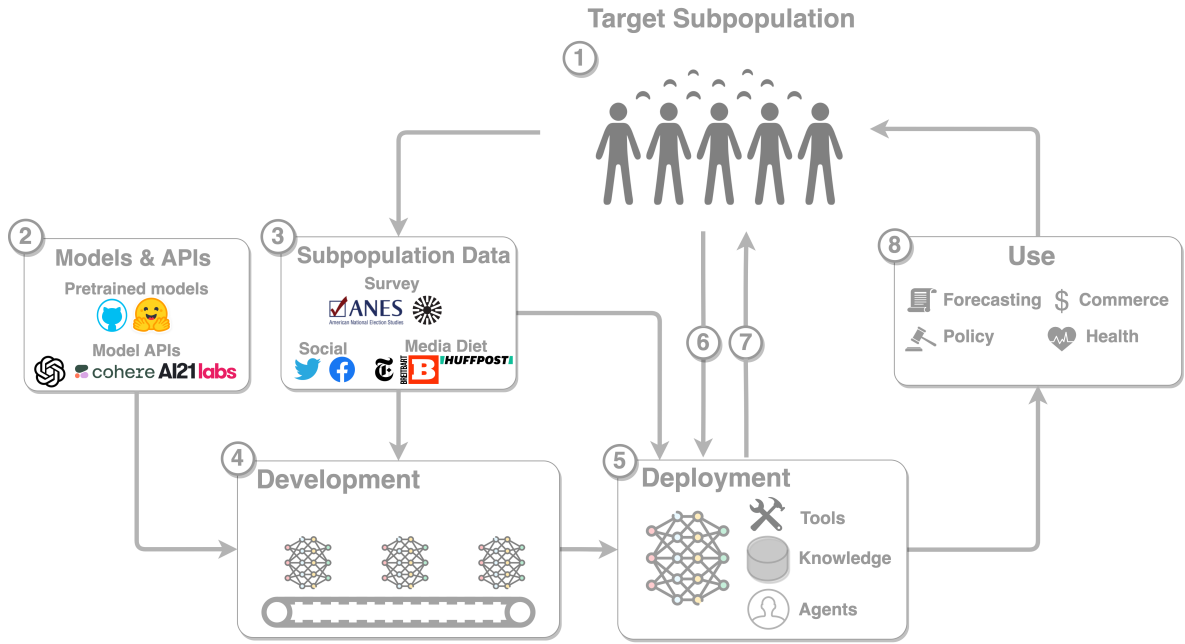

Figure 1: Subpopulation representative models (SRMs) represent a target subpopulation ⓵ using a large language model (LLM). SRMs may elicit representative behavior by prompting or fine-tuning existing pre-trained or fine-tuned models, either locally or via a cloud API ⓶. Representative behavior may be grounded with pre-existing data from the target subpopulation (potentially repurposed from social and traditional media or survey data), or with bespoke data ⓷. During the model development process, variables are selected, models are tuned to elicit subpopulation representative behavior ⓸. At deployment ⓹, SRMs may be augmented with additional data or tools. In public-facing deployments, target and non-target subpopulations may interact with the deployed model directly ⓺ (i.e., to give feedback on model outputs). SRMs may directly interact with the target subpopulation or other populations not involved in model development ⓻ (e.g., through a dialog interface like ChatGPT) or have indirect effects on target or non-target subpopulations when used for decision-making in domains including policy, forecasting, and commerce applications ⓼.

defining it as "the degree to which the complex patterns of relationships between ideas, attitudes, and socio-cultural contexts within a model accurately mirror those within a range of human subpopulations". It is difficult to establish a clear threshold for what makes a model's behavior complex, and it is expected that an approximation of simple phenomena may also be useful in some settings. Argyle's algorithmic fidelity points to a particular range on the spectrum of SRM task complexity which we introduce in Section 5.3.1.

Our definition of SRB can be seen as a relaxation of the concept of algorithmic fidelity to cover all levels of complexity. Santurkar et al. (2023) proposed the concept of *opinion alignment*, defining it mathematically as "an average distance between the distribution of SRM outputs and the distribution of target subpopulation responses over a set of questions with categorical answer types [i.e., a Likert scale or multiple-choice question]." This definition lends itself specifically to empirical study of SRM behavior in closed-ended tasks. However, many of the surveyed works (Section 5) apply SRMs for open-ended tasks. Our definitions are intentionally broad, allowing us to gather a diverse set of SRM studies based on their important commonal-ities, namely the use of large language models to represent subpopulation behavior.

Figure 1 provides a general overview of the processes involved in using LLMs to estimate SRMs. Most of the specific implementations we review in the next section are concerned with steps in the front-end of the workflow (e.g., Stages 1-4), but we also discuss issues and concerns involved in the later stages (6-8) in Section 5.

## 4. Subpopulation Representative Model Implementations

This section provides a review of present applications of large language models to estimate subpopulation representative models. Table 1 summarizes the SRM system, the LLM used to implement it, the training paradigm, the communities represented by each SRM, and the tasks for which each SRM are intended.

### 4.1. SRM Domains

Subpopulation representative models have been studied or proposed for use in a variety of domains. These include political tasks including predicting election outcomes, gauging the popularity of political parties and candidates, and understanding voter preferences on specific policy issues.[43] This is closely connected, though not identical, to the use of SRMs for sociological tasks[44] or for prototyping experiments in sociology and psychology.[45]

Proposed use cases also include commercial applications—gathering consumer opinions to improve products, identify market gaps, develop targeted marketing campaigns, track brand perception, and identify opportunities for brand positioning.[46] SRMs have also been proposed as tools in the public health domain[47] and to inform policy.[48]

The projects we surveyed vary in tone, from critical to optimistic. The association of a project with a particular domain does not necessarily imply that the authors proposed SRM use in the domain, rather that this was the domain in which the SRM was proposed for use or evaluated.

### 4.2. Target Subpopulations

Most of the SRM implementations and proposals we identified target subpopulations in the United States. Exceptions include the system discussed in Palakodety, KhudaBukhsh, and Carbonell (2020) , which was used to analyze the 2019 Indian General Election, and the Romanian government's proposed ION system (*ION - Primul Consilier Cu Inteligență Artificială Al Guvernului* 2023).

[43] Eric Chu et al., *Language Models Trained on Media Diets Can Predict Public Opinion*, arXiv, 2023, https://doi.org/10.48550/arXiv.2303.16779., Hang Jiang et al., "CommunityLM: Probing Partisan Worldviews from Language Models," in *Proceedings of the 29th International Conference On Computational Linguistics* (Gyeongju, Republic of Korea: International Committee on Computational Linguistics, 2022), 6818–6826, https://aclanthology.org/2022.coling-1.593., Shriphani Palakodety, Ashiqur R. KhudaBukhsh, and J. Carbonell, *Mining Insights from Large-Scale Corpora Using Fine-Tuned Language Models*, Amsterdam: IOS Press, 2020, 1890–1897., Lisa P. Argyle et al., "Out of One, Many: Using Language Models to Simulate Human Samples," *Political Analysis* 31, no. 3 (2023): 337–351.

[44] Philip Feldman et al., *Polling Latent Opinions: A Method for Computational Sociolinguistics Using Transformer Language Models*, arXiv, 2022, https://doi.org/10.48550/arXiv.2204.07483., Philip G. Feldman, Shimei Pan, and James Foulds, "The Keyword Explorer Suite: A Toolkit for Understanding Online Populations," in *Companion Proceedings of the 28th International Conference on Intelligent User Interfaces* (New York: Association for Computing Machinery, 2023), 21–24, https://doi.org/10.1145/3581754.3584122. , Junsol Kim and Byungkyu Lee, *AI-Augmented Surveys: Leveraging Large Language Models for Opinion Prediction in Nationally Representative Surveys*, arXiv, 2023, https://doi.org/10.48550/arXiv.2305.09620.

[45] Gati Aher, Rosa I. Arriaga, and Adam Tauman Kalai, *Using Large Language Models to Simulate Multiple Humans and Replicate Human Subject Studies*, arXiv, 2023, https://doi.org/10.48550/arXiv.2208.10264.

[46] Chu et al., *Language Models Trained on Media Diets Can Predict Public Opinion.*

[47] Chu et al.

[48] Agence France-Presse, "Romania PM Unveils AI 'adviser' to Tell Him What People Think in Real Time," *The Guardian*, March 2023. https://www.theguardian.com/world/2023/mar/02/romania-ion-ai-government-honorary-adviser-artificial-intelligence-pm-nicolae-ciuca.

Table 1: Implementations of subpopulation representative models. **Elicitation techniques** FT: Fine-Tuning; P: Prompting; D: Distillation; PEFT: Parameter-Efficient Fine-Tuning; RFT: Reward Model Fine-Tuning. **Inference techniques** OTG: Open-ended Text Generation; CTG: Close/Closed-ended Text Generation; CLS: Classification; C: Clustering; SA: Sentiment Analysis; D: Dialog; Unk: Unknown. **Risks identified** HTG: Harmful Text Generation; Misinfo: Misinformation; Fid: Poor Fidelity; Bias: Model Bias or Lack of Fairness; Perf: General performance issues; Privacy: Potential to reveal sensitive information.

| Method | Models | Elicitation techniques | Inference techniques | Tasks | Domains | Data sources | Subpopulations | Risks identified |
|---|---|---|---|---|---|---|---|---|
| CommunityLM (H. Jiang et al. 2022) | GPT-2 | FT | OTG, SA | Opinion mining | Politics | Twitter | American partisan groups | Erasure |
| Silicon sampling (Argyle et al. 2023) | GPT-3 | P | OTG, CTG | Behavior prediction, opinion mining | Politics, sociology | Public opinion surveys | American ideological groups, multiple | Misuse, Misinfo |
| Media diet models (Chu et al. 2023) | BERT | FT, P | CTG | Opinion mining, forecasting, media effects analysis | Public health, commerce | Media outlets, public opinion surveys | American media consumers | Misuse, Fid, Bias, Perf, Misinfo |
| The Keyword Explorer Suite (P.G. Feldman et al. 2023) | GPT-2, GPT-3 | FT | OTG, SA | Forecasting, opinion mining | Sociology | Social media | Multiple | Misuse |
| Polling latent opinions (P. Feldman et al. 2022) | GPT-2 | FT | OTG, SA | Opinion mining | Commerce | Yelp reviews | Restaurant customer subgroups | HTG, Bias Misinfo, Fid, Misuse, Perf |
| Turing experiment (Aher, Arriaga, and Kalai 2023) | GPT-3 | P | OTG | Behavior prediction | Psychology | Human psychology experiments | Psychology study participants | Bias, HTG, Privacy |
| Palakodety, KhudaBukhsh, and Carbonell (2020) | BERT | FT | CTG | Opinion mining | Politics, policy | Youtube comments | Election commenters on YouTube | Qual |
| Bisbee et al. (2023) | ChatGPT-3.5-turbo | P | CTG | Opinion mining | Politics, sociology | Public opinion surveys | American ideological groups, multiple | Bias |
| AI-augmented surveys (Kim and B. Lee 2023) | DCN + Alpaca-7B | FT | CLS | Opinion mining | Sociology | Public opinion surveys | US citizens | Bias, Misuse, Privacy |
| SFT-Utilitarian (M. A. Bakker et al. 2022) | Chinchilla | P, RFT | OTG | Opinion mining, consensus generation | Policy | Bespoke data, crowd-sourced annotation | UK citizens | Misinfo, Fid, Bias, Misuse |
| Moral Mimicry (Simmons 2023) | GPT-3, OPT | P | OTG | Opinion mining | Psychology | Crowd-sourced data, online forums | American ideological groups | HTG, Misinfo |
| ION (*ION*, n.d.) | Unk | Unk | OTG | Opinion mining | Policy | User messages, Unk | Romanian citizenry | Unk |
| Talk to the City (TtC) | OpenAI, Anthropic | FT | C, D | Opinion mining | Policy | Surveys, speeches, posts, Unk | City residents | Bias, Fid, Misinfo |

## 4.3. Tasks

This section summarizes the tasks that existing SRM systems have been used for. We can roughly divide existing tasks into four categories: *opinion mining*, *behavior prediction*, *forecasting*, and *sociological analysis*. Examples in each category are presented in this section. We note that these categories are not derived from a formal taxonomy and are not necessarily mutually exclusive. Additionally, future SRMs may be used for tasks that do not fall squarely into one of the categories below.

### 4.3.1. Task domains

1. **Opinion mining**: Defined as "extracting opinions from unstructured texts through the combination of NLP and data science techniques."[49] Opinion mining was the most popular task across the papers we surveyed. Argyle et al. (2023) mine partisan opinions about shared and opposing political affiliations using open-ended text generation, by prompting GPT-3 with partisan demographic data from the Pigeonholing Partisans dataset.[50] H. Jiang et al. (2022) extract partisan opinions about political figures and social groups from ANES Surveys using GPT-2 models finetuned on Twitter data. (Chu et al. 2023) use BERT models fine-tuned on online news articles, TV transcripts, and radio show transcripts to synthesize media diet consumer opinions on public health and consumer confidence. (M. A. Bakker et al. 2022) apply prompting and reward model fine-tuning to Chinchilla 70B to generate consensus opinions from crowdsourced data. Additionally, governance projects propose to use LLMs to mine and aggregate opinions at the municipal[51] and national[52] level.

2. **Behavior prediction**: For the scope of this work, we define behavior prediction as any task "attempting to predict the behavior of a prototypical member of the modeled subpopulation". We note that this definition blurs the line between modeling subpopulations and modeling individuals. However, we argue that use of LLMs for behavior prediction ex-tends existing work in the social sciences that uses individual-level predictive modeling to investigate macroscopic (subpopulation) level phenomena.[53] Argyle et al. (2023) use GPT-3 for vote prediction, by prompting the model with partisan demographic data to generate political candidate names. Aher, Arriaga, and Kalai (2023) attempt to replicate the results of seminal human psychology experiments using GPT-3 as a simulator.

3. **Forecasting**: Behavior prediction is closely related to forecasting[54], where the goal is to "use a subpopulation representative model to predict future events or trends". Chu et al. (2023) proposed the use of fine-tuned BERT models to forecast public sentiment in response to events, using the COVID-19 pandemic as an example.

4. **Sociological analysis**: Here, the task is to use an SRM to gain insights into the target subpopulation to enhance scientific understanding. Chu et al. (2023) exemplifies this approach, demonstrating how media diet models can be used to understand the effect of media consumption on response variables such as consumer confidence. Chu et al. (2023) explores whether the moral attitudes of ideological groups (self-identified liberals and conservatives) are replicated in the outputs of GPT and Open Pre-Trained (OPT) LLMs.

[49] Chaima Messaoudi, Zahia Guessoum, and Lotfi Ben Romdhane, "Opinion Mining in Online Social Media: A Survey," *Social Network Analysis and Mining* 12, no. 1 (2022): 25. https://doi.org/10.1007/s13278-021-00855-8.

[50] Jacob E. Rothschild et al., "Pigeonholing Partisans: Stereotypes of Party Supporters and Partisan Polarization," *Political Behavior* 41, no. 2 (2019): 423–443. https://doi.org/10.1007/s11109-018-9457-5.

[51] AI Objectives Institute, *Introducing Talk To the City: Collective Deliberation at Scale*, 2023, https://www.talktothe.city/.

[52] *ION - Primul Consilier Cu Inteligență Artificială Al Guvernului*, 2023, https://ion.gov.ro/.

[53] Scott De Marchi and Scott E. Page, "Agent-Based Models," *Annual Review of Political Science* 17 (2014): 1–20.

[54] Fotios Petropoulos et al., "Forecasting: Theory and Practice," *International Journal of Forecasting* 38, no. 3 (2022): 705–871.

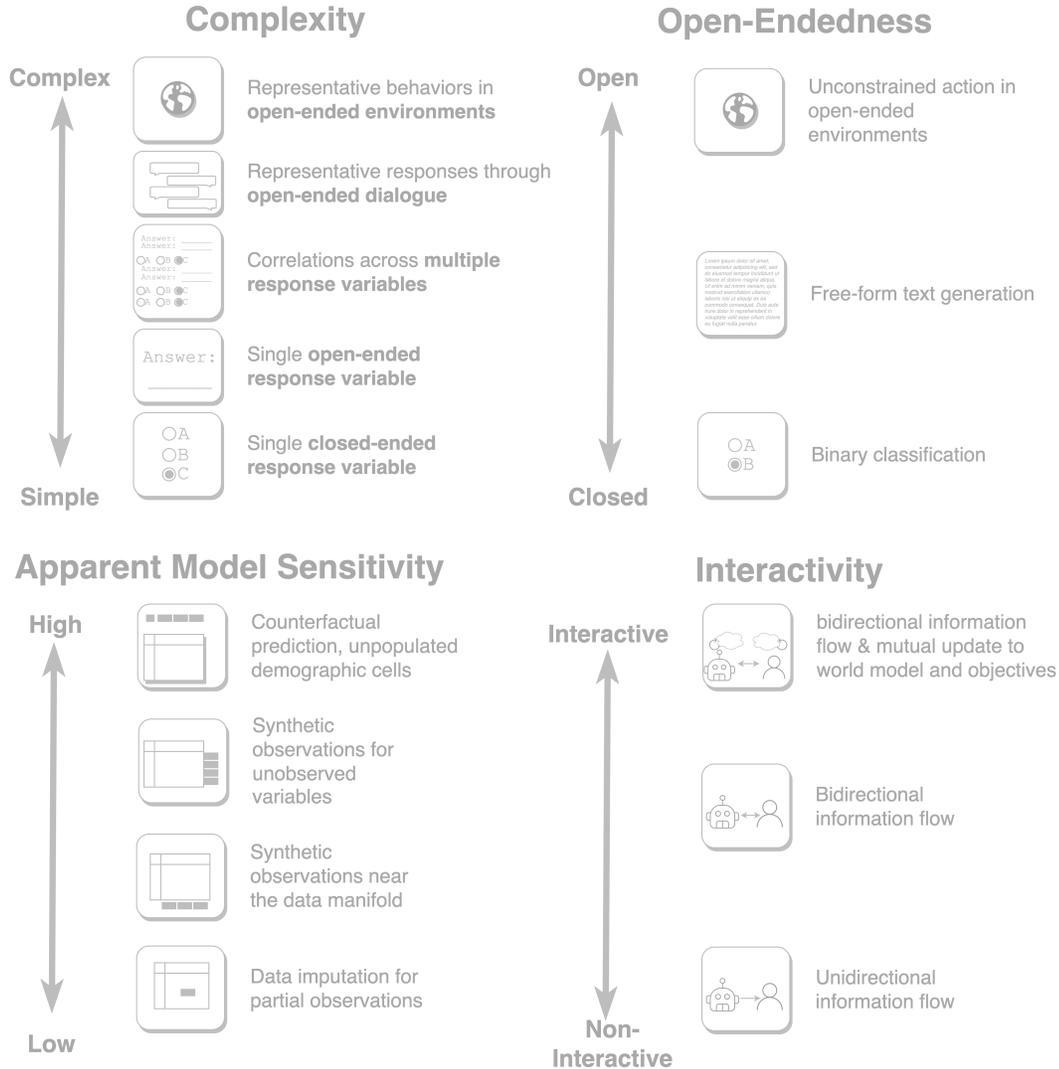

Figure 2: Task complexity spectrum for subpopulation representative models. Subpopulation representative modeling varies in complexity from modeling the response distribution over single-closed and open-ended output variables, to modeling the joint conditional response distribution over multiple output variables, to modeling the output distribution over dynamic output variables (e.g., determined through dialog), to planning and executing representative behavior in an open-ended environment.

### 4.3.2. TASK DIMENSIONS

In addition to a domain-based classification of subpopulation representative modeling tasks, we also propose to characterize subpopulation representative modeling tasks across several task dimensions (illustrated in Figure 2). We consider this approach useful for a number of reasons. First, a multidimensional characterization of subpopulation representative modeling provides a frame to compare SRM tasks and approaches. This may be useful for direct comparison between SRM implementations, as well as to detect trends in SRM implementations over time. For autonomous vehicles, classifications of system capability aid effective development, use, and regulation of the technology[55] frames for classifying SRM systems

[55]Girish Chowdhary, Chinmay Soman, and Katherine Driggs-Campbell, *Levels of Autonomy for Field Robots*,

could yield similar benefits. Finally, each of the task dimensions we identified highlights an opportunity for interdisciplinary work.

In the remainder of this section, we discuss four salient dimensions to characterize SRM tasks: *complexity*, *open-endedness*, *model sensitivity*, and *interactivity* (see Figure 2). For each of these dimensions, we provide our working definition, relations to existing definitions, examples of SRMs at various levels, and the significance of the dimension to SRM research. We recognize that this taxonomy is far from final, and welcome further efforts to clarify these concepts, identify additional relevant SRM task dimensions, and operationalize these dimensions for formal study.

1. **Apparent task complexity**

   - *Definition.* Apparent task complexity refers to the "level of behavioral or attitudinal sophistication that a subpopulation representative model is tasked to represent". SRM tasks aim to replicate human behaviors, but (as we discuss below) achieving similar outputs between humans and LLMs does not imply underlying mechanisms of equal complexity. By apparent task complexity, we refer to how complex a particular task seems in humans, cautioning that this apparent complexity is likely correlated but not necessarily equal to the complexity of LLM behavior required to succeed at the task.

   - *Relation to other definitions.* Various disciplines define complexity differently: computer science focuses on computational resources[56]; complex systems theory on component diversity and interactions[57]; management and behavioral sciences on nested compositions of behaviors[58]; information theory on sequence predictability[59]; and Kolmogorov complexity on minimal program lengths.[60] Meanwhile, a universal definition

of task complexity in machine learning remains an active research topic.[61]

In political science, the most influential models of public opinion[62] draw on an information-theoretic sense of complexity, emphasizing unpredictability as the defining complexity measure of individual-level political attitudes.[63] In this understanding, attitudes are mixtures of competing considerations relevant to the attitude object and survey responses represent a sample of the most salient considerations. From this perspective, unpredictability can result from either "ambivalence" (when considerations are present but cross-cutting) or "indifference" (when considerations are absent) (e.g., Thornton, 2011). Both ambivalence and indifference vary systematically across individuals and objects. Namely, attitude certainty is positively correlated with political sophistication, partisan attachments, and personality traits like rigidity and the need for closure.[64] Certainty also varies at the item-level, and is generally greater for less technical and more familiar issues.[65]

   - *Placing SRMs on the complexity spectrum.* The studies surveyed here already span a range of complexity levels, from simple multiple-choice classification tasks (e.g., vote prediction) to correlations across multiple response variables (e.g., *pattern correspondence* in Argyle et al., 2023).

We can imagine this task complexity spectrum extending further—the recent popularity of ChatGPT illustrates the viability of interactive LLM-based systems. In the equivalent SRM system, this would require representative behavior spanning multiple turns in a dialog setting. Systems like ION and Talk to the City hint at this level of capability.[66] The complexity spectrum may go further still. One of the frontier areas of development in LLM technology is the use of LLM agents (see Section 2.3). This direction is hinted at by J. S. Park et al. (2023) although the authors don't design their system to mimic a human subpopulation, it is not difficult to imagine a similar multi-agent system designed to mimic a specific human subpopulation.

- *Why is it important to consider apparent task complexity?* Recent empirical results in natural language processing suggest that more complex behavior often requires larger models, or conversely that larger models often display superior abilities. Knowledge of an approximate complexity level for a task is likely to aid design of an appropriately-sized SRM system. Behavior complexity may also correlate with the amount or kind of data required to control model sensitivity in the SRM system, informing data collection requirements during system development. Additionally, system complexity informs the choice of evaluation techniques, with more complex intended system behavior often requiring more complex evaluation.

In political science, an understanding of attitudinal complexity helps to explain the regularity of phenomena such as response instability and wording/order effects in public opinion survey research. If attitudes are heterogeneous mixtures of considerations rather than fixed quantities, survey responses will be sensitive even in the absence of other kinds of measurement error. Accordingly, if opinions are indeed better understood as "fuzzy" distributions rather than fixed quantities, it is crucial that LLMs are assessed on the distributional properties of their simulated responses. Indeed, because LLMs allow us to repeatedly sample from attitudinal distributions by running multiple simulations of a given prompt, they afford a unique opportunity to quantify our uncertainty about individual opinions in a way that does not require multi-wave panel data (e.g., Alwin and Krosnick, 1991).

## 2. **Open-endedness**

[56]Walter Dean, "Computational Complexity Theory," in *The Stanford Encyclopedia Of Philosophy*, ed. Edward N. Zalta, Metaphysics Research Lab, Stanford University, 2021, https://plato.stanford.edu/archives/fall2021/entries/computational-complexity/.

[57]Christopher Magee and Olivier de Weck, *Complex System Classification* (International Council On Systems Engineering (INCOSE), 2004), accessed July 22, 2023, https://dspace.mit.edu/handle/1721.1/6753.

[58]Michael Lamport Commons, "Introduction to the Model of Hierarchical Complexity," *Behavioral Development Bulletin* 13, no. 1 (2007): 1–6. https://doi.org/10.1037/h0100493.

[59]John Hale, "Information-Theoretical Complexity Metrics," *Language and Linguistics Compass* 10, no. 9 (2016): 397–412. https://doi.org/10.1111/lnc3.12196.

[60]Ming Li and Paul Vitányi, *An Introduction to Kolmogorov Compelxity and Its Applications*, 4th ed., Texts in Computer Science (New York: Springer, 2019).

[61]Alessandro Achille et al., *The Information Complexity of Learning Tasks, Their Structure and Their Distance*, arXiv, 2020, https://doi.org/10.48550/arXiv.1904.03292., Akhilan Boopathy et al., "Model-Agnostic Measure of Generalization Difficulty," in *Proceedings of the 40th International Conference on Machine Learning*, eds. Andreas Krause et al. (PMLR, 2023), 2857–2884, https://proceedings.mlr.press/v202/boopathy23a.html.

[62]Philip E. Converse, "The Nature of Belief Systems in Mass Publics," in *Ideology and Discontent*, ed. David E. Apter, New York: Free Press, 1964, 206–261., John Zaller, *The Nature and Origins of Mass Opinion* (New York: Cambridge University Press, 1992)., John Zaller and Stanley Feldman, "A Simple Theory of the Survey Response: Answering Questions Versus Revealing Preferences," *American Journal of Political Science* 36, no. 3 (1992): 579–616.

[63]See also Alvarez and Brehm (2002) .

[64]John T. Jost et al., "Political Conservatism as Motivated Social Cognition," *Psychological Bulletin* 129, no. 3 (2003): 339–375.

[65]Charles S. Taber and Lodge Milton, "Motivated Skepticism in the Evaluation of Political Beliefs," *American Journal of Political Science* 50, no. 3 (2006): 755–769., Dan M. Kahan, "Ideology, Motivated Reasoning, and Cognitive Reflection," *Judgment and Decision Making* 8, no. 4 (2013): 407–424. , Justin H Gross and Daniel Manrique-Vallier, "A Mixed Membership Approach to Political Ideology," in *Handbook of Mixed Membership Models and Their Applications*, eds. Edoardo M. Airoldi et al., Boca Raton, FL: CRC Press, 2015, 119–140., Edward G. Carmines and James A. Stimson, "The Two Faces of Issue Voting," *American Political Science Review* 74, no. 1 (1980): 78–91.

[66]AI Objectives Institute, *Introducing Talk To the City: Collective Deliberation at Scale*, 2023, https://www.talktothe.city/., ION - Primul Consilier Cu Inteligență Artificială Al Guvernului, 2023, https://ion.gov.ro/.

- *Definition.* Open-endedness refers to "the extent to which SRM system behavior is unconstrained". A closed-ended system is one whose outputs are constrained to some narrow range imposed by the system developer.

- *Placing SRMs on the open-endedness spectrum.* For the most part, the surveyed studies display a low degree of open-endedness ranging from binary classification to open-ended response generation. Several authors frame vote prediction as a classification task, restricting the model to generate outputs from a limited set of candidates.[67] Aher, Arriaga, and Kalai (2023) restrict their simulations of psychological survey participants to a list of valid actions in the simulated experimental setup. The "partisan descriptors" task proposed by Argyle et al. (2023) is more open-ended—the SRM system can use any word from the model vocabulary as a descriptor, and the system response is only limited by length. We expect that future SRM systems will trend towards more open-ended tasks. These might include long, open-ended text responses, or even interactive dialogues with simulated subpopulation members to understand political issue saliency. The open-endedness spectrum may extend even further to LLM-based agents taking open-ended actions in simulated or real environments.

- *Why is it important to consider task open-endedness?* Considering the level of open-endedness of an SRM system is important for several reasons. Many of the risks posed by LLM systems and algorithmic systems in general[68] are potentiated by the system being able to produce unconstrained outputs. Additionally, the open-endedness of an SRM system also informs the techniques that will be required to evaluate it. While constrained tasks like regression and classification have a host of performance metrics familiar to statisticians and machine learning practitioners, evaluating open-ended language is less clear. Natural language processing has long wrestled with the difficulty of defining metrics for open-ended tasks like summarization, developing methods like BLEU[69] and ROUGE[70]. Recently, there has been a trend towards model-based evaluation, earlier examples include BERTScore[71] and recently larger models have been used to form more complex evaluations beyond textual similarity.[72] The challenge of reliably evaluating SRM systems without restricting their use to closed-ended settings leads directly into the problem of scalable oversight.[73]

3. **Model sensitivity**

- *Definition.* Model sensitivity refers to "the extent to which the SRM task relies on the

[67] Lisa P. Argyle et al., "Out of One, Many: Using Language Models to Simulate Human Samples," *Political Analysis* 31, no. 3 (2023): 337–351., Shriphani Palakodety, Ashiqur R. KhudaBukhsh, and J. Carbonell, *Mining Insights from Large-Scale Corpora Using Fine-Tuned Language Models*, Amsterdam: IOS Press, 2020, 1890–1897.

[68] Renee Shelby et al., *Identifying Sociotechnical Harms of Algorithmic Systems: Scoping a Taxonomy for Harm Reduction*, arXiv, 2023, https://doi.org/10.48550/arXiv.2210.05791. , Laura Weidinger et al., "Taxonomy of Risks Posed by Language Models," in *2022 ACM Conference On Fairness, Accountability, And Transparency* (New York: Association for Computing Machinery, 2022), 214–229, https://doi.org/10.1145/3531146.3533088.

[69] Kishore Papineni et al., "BLEU: A Method for Automatic Evaluation of Machine Translation," in *Proceedings of the 40th Annual Meeting Of the Association For Computational Linguistics* (Philadelphia, PA: Association for Computational Linguistics, 2002), 311–318, https://doi.org/10.3115/1073083.1073135.

[70] Chin-Yew Lin, "ROUGE: A Package For Automatic Evaluation Of Summaries," in *Text Summarization Branches Out* (Barcelona, Spain: Association for Computational Linguistics, 2004), 74–81, https://aclanthology.org/W04-1013.

[71] Tianyi Zhang et al., *BERTScore: Evaluating Text Generation with BERT*, arXiv, 2020, https://doi.org/10.48550/arXiv.1904.09675.

[72] Ethan Perez et al., *Discovering Language Model Behaviors with Model-Written Evaluations*, arXiv, 2022, https://doi.org/10.48550/arXiv.2212.09251.

[73] Dario Amodei et al., *Concrete Problems In AI Safety*, arXiv, 2016, https://doi.org/10.48550/arXiv.1606.06565., Samuel R. Bowman et al., *Measuring Progress on Scalable Oversight for Large Language Models*, arXiv, 2022, https://doi.org/10.48550/arXiv.2211.03540.

extrapolative capabilities of the LLM", as opposed to relying directly on observed data.

- *Relation to other definitions*. We find it useful to distinguish between two notions of model sensitivity—apparent and actual. *Apparent model sensitivity* is what the practitioner might expect from the learning setup, based on how much task-relevant information is available in the training data, and the empirical generalization properties of the model class. *Actual model sensitivity* is the real extent to which a prediction on some input data depends on the model structure. These two notions often agree in practice. For example, few-shot learning often achieves higher performance than zero-shot learning, all other factors being held equal. However, our intuitions are not always correct - adversarial examples are inputs for which the model "fails" despite our expectations that it should be able to generalize.[74]

Theories of learning aim to bridge the gap between our expectations and real model performance. Traditional approaches for predicting model sensitivity were developed in an era when machine learning was most often used on tabular data. Some approaches calculate whether an input example appears within the convex hull of the training data.[75] King and Zeng (2006) proposed to instead calculate a diameter across the data manifold, and compare this dataset diameter to the distance between the training data and a new sample that we would like to predict on. It is important to note that both of these approaches rely on a geometrical intuition: models perform better for examples that are closer to the data used for training. These methods rely on being able to calculate meaningful distances between data points in a model-agnostic way. However, deep learning excels specifically in settings where this is challenging. Representation learning—the ability for networks to learn meaningful encodings for high-dimensional input data—is often touted as a key strength of deep learning. The embedding space learned by the model does allow for the calculation of meaningful pointwise distances, but these are inherently model-dependent.

Much theoretical work has been done in an attempt to make model-agnostic predictions about generalization in these more challenging settings. Statistical learning theory formalizes the conditions required for generalization in classical machine learning models[76], but fails to explain the success of deep neural networks. Singular learning theory aims to resolve the deficiencies of statistical learning theory in explaining highly expressive models like DNNs[77], but computational challenges limit its direct applicability to the large networks used in practice. Despite the shortcomings of formal learning theory, our intuitive expectations about model sensitivity *do* often hold up in practice. Therefore, we suggest practitioners consider the apparent model sensitivity of their task, knowing that their expectations are not guaranteed. We comment on the apparent model sensitivity of some of the surveyed SRMs below.

However, due to the differences between traditional learning and deep learning, it is likely that these approaches may not be sufficient to accurately predict model sensitivity in the SRM setting. Namely, it is unclear how one would measure "distance" in a model agnostic way in tandem with an LLM: the (textual) data itself is sparse and high-dimensional (and hence, not amenable to distance or similarity measures); while the dense and lower-dimensional embedding space is a product of the model itself.

- *Placing SRMs on the apparent model sensitivity spectrum*. We expect model sensitivity to be low when the model has plentiful data that is similar to the task examples at hand for all relevant notions of similarity. These notions of similarity can include factors of the predictive task including the demographic variables, response variables, as well as other factors. (Kim and B. Lee 2023) provide examples of SRM tasks at varying levels of apparent model sensitivity: missing data imputation, retrodiction, and zero-shot prediction. Missing data imputation—where the model is tasked with predicting the value of missing variables for partial observations—illustrates the least apparent model sensitivity. In this setting, the model has access to values of the target variable for other observations, and values of other variables for the target observation. In the retrodiction setting, the model does not have access to the target variable for the time period in question, but has access to target variable values for other years. The zero-shot prediction setting displays the greatest degree of model sensitivity, as the model has no access to observed values of the target variable.

- *Why is model sensitivity important?* Model sensitivity is closely related to the suitability of a model for counterfactual prediction[78], and the ability to predict outcomes in counterfactual scenarios is of fundamental importance to decision-making. Furthermore, prediction for sparsely populated cells is of particular interest in political analysis settings.[79] While actual model sensitivity sometimes diverges from our expectations, the basic intuition that models will do better with more task-relevant data often holds up in practice. Therefore, thinking about apparent model sensitivity is not wasted effort. From the model developer's perspective, model sensitivity allows us to put useful bounds on our expectations about model performance. Knowledge about general relationships between the learning setup and generalization also allow us to estimate the quality and quantity of data required when developing a new SRM system.

From a regulatory perspective, a precise understanding of model generalization may allow more precise regulation. Knowledge that certain learning settings exhibit greater model sensitivity than other settings could allow targeted allocation of regulatory effort towards these settings. Additionally, knowing that a certain learning setup exhibits low model sensitivity allows allocation of regulatory effort towards the data. From the perspective of users who will interact directly with SRMs, or members of the subpopulation affected by SRM performance, model sensitivity is a key ingredient of trust.

## 4. Interactivity

- *Definition.* We define interactivity as the "degree of information sharing and mutual influence between an SRM system and a user of the system". In gauging the interaction between humans and LLM-based systems, we suggest considering the following four questions:

[74]Ian J. Goodfellow, Jonathon Shlens, and Christian Szegedy, "Explaining and Harnessing Adversarial Examples," in *3rd International Conference On Learning Representations, ICLR 2015, San Diego, CA, USA, May 7-9, 2015, Conference Track Proceedings*, eds. Yoshua Bengio and Yann LeCun (2015), https://doi.org/10.48550/arXiv.1412.6572.

[75]Trevor Hastie, Robert Tibshirani, and Jerome Friedman, *The Elements of Statistical Learning*, 2nd ed., Springer Series in Statistics (New York: Springer, 2009).

[76]V.N. Vapnik, "An Overview of Statistical Learning Theory," *IEEE Transactions on Neural Networks* 10, no. 5 (1999): 988–999. https://doi.org/10.1109/72.788640.

[77]Sumio Watanabe, "Singular Learning Theory," in *Algebraic Geometry and Statistical Learning Theory*, ed. Sumio Watanabe, Cambridge Monographs on Applied and Computational Mathematics, New York: Cambridge University Press, 2009, 158–216, https://doi.org/10.1017/CBO9780511800474.007.

[78]Gary King and Langche Zeng, "The Dangers of Extreme Counterfactuals," *Political Analysis* 14, no. 2 (2006): 131–159. https://doi.org/10.1093/pan/mpj004.

[79]Challenges of counterfactual prediction and the sparse cell problem are explored in Section 6.4.

(a) How much *information* is shared? How often does interaction occur (in absolute terms and in relation to the frequency of decision-making)?

(b) To what extent is the *world model* of each interactant updated?

(c) To what extent are the *objectives* of each party updated?

(d) To what extent are each of these communications and influences *bidirectional*?[80]

- *Factors of interactivity.* In common usage, interactivity refers to communication and mutual influence that occurs between entities, such as humans, organizations, or computer programs. Definitions from the field of media studies focus on the extent of information exchange, the responsiveness of a system to a user, and the degree of control the user has over system operation.[81] Z. Lin et al. (2023) proposed a design space for mixed-initiative co-creative systems with three dimensions: human-initiated vs. AI-initiated, reflection vs. elaboration, and local vs. global. This framework is intended to describe systems where AI and human counterparts collaborate on a creative artifact such as storytelling. Lin et al. refer to prior work by Novick and Sutton (1997) who defines a mixed-initiative system as a system where human and computational initiatives cooperate towards a shared goal.

- *Placing SRMs on the interactivity spectrum.* Most uses for machine learning to date lie at the low end of the interactivity spectrum, including most of the SRMs we surveyed. This includes tasks like vote prediction[82] and partisan description, where the behavior of the model can be performed in a single LLM inference. At the high end of the interactivity spectrum, we can imagine a task where an LLM-based system and a human operator frequently share information with each other, updating each other's world model, and updating each other's objectives in some open-ended pursuit. Project proposals such as ION and Talk to the City are examples of SRM systems that aim to occupy a higher position on the interactivity spectrum. Both projects propose LLM-based systems where governments (at the national and city levels, respectively) can learn about their constituents through interaction with the system. Based on these proposals, it appears that most of the flow of information is in the direction from the LLM system to the user. LLM "agents" are another LLM system design pattern where users set high-level objectives that the LLM system pursues autonomously.[83] This pattern has not yet appeared in the SRM domain. The greatest degree of interactivity would involve an LLM system and a human user mutually sharing information resulting in meaningful changes to goals and world models of the other, much like the interaction between political representatives and their constituents in today's democracy.

- *Why is it important to consider interactivity?* Governance processes are inherently interactive, but the speed and scope of interaction is often constrained by the difficulty for representatives to interact with a large, geographically distributed polity. Thus, it is reasonable to expect interest in technologies that make this process scalable, so it is important to understand a need for activity as a driving factor for SRM development. As we describe in Section 5.3, new technological capabilities tend to lead to new risks - interactivity is no exception. As an example, systems that allow or encourage users to share private information introduce security concerns that are not present for systems where users simply obtain information. Like open-endedness, the interactivity level of a system also informs the kind of evaluation that will be required, with greater degrees of interactivity posing greater challenges for evaluation. While measures for fidelity and alignment in non-interactive settings are already being developed[84], evaluations for fidelity in interactive systems are not yet well-studied. With growing interest around LLM dialogue systems and agents, we expect that evaluations of fidelity in more interactive and agentic SRMs settings will be an emerging research area with potential for contribution from multiple disciplines.

- **Other dimensions**

  In addition to complexity, open-endedness, sensitivity, and interactivity, other dimensions to characterize SRM tasks may include *task temporality* (the temporal relationship between SRM training data and predictions) and *stereotyping* (the extent to which SRMs model individuals vs. prototypical members of the subpopulation). We omit these dimensions here for brevity, but note that they are also important ways to characterize SRM tasks.

### 4.4. MODEL ARCHITECTURES AND ELICITATION TECHNIQUES

The studies we surveyed use a variety of language models. CommunityLM (H. Jiang et al., 2022), "Polling Latent Opinions" (P. Feldman et al., 2022), and "Keyword Explorer Suite" (P. G. Feldman, Pan, Shimei, and J. Foulds, 2023) use fine-tuned GPT-2[85] models. Chu et al. (2023) and Palakodety, KhudaBukhsh, and Carbonell (2020) use fine-tuned BERT[86] models. Aher's Turing experiment[87], and Argyle's silicon ampling technique[88] apply prompting techniques to OpenAI's GPT-3 model, and Bisbee et al. (2023) applies prompting to ChatGPT (*gpt-3.5-turbo*). The AI Objectives Institute's Talk to the City proposal[89] does not disclose specific models, but lists "OpenAI and Anthropic LLMs." M. A. Bakker et al. (2022) applies prompting and reward model fine-tuning to the Chinchilla 70B[90] model. Simmons (2023) prompts GPT-3 and OPT models to simulate subpopulation moral preferences.

Kim and B. Lee (2023) fine-tune a Deep Cross Network (DCN) architecture consisting of a language model to embed survey questions (`Alpaca-7B`[91] and `GPT-J-6B`[92]) were used, and individual and temporal embeddings to encode individual beliefs across questions and the survey period. The DCN uses the embeddings to produce a binary response to the survey question, framing survey response prediction as classification.

[80]In considering questions of objectives and bidirectional communication, we note that the LLM system, by itself, lacks agency. However, it is possible to instantiate (via an LLM) a simulacra with apparent objectives, and it is possible to update these via apparent interaction between the user and the simulacra.

[81]John December, "Units of Analysis for Internet Communication," *Journal of Communication* 46, no. 1 (1996): 14–38., John Pavlik, *New Media Technology: Cultural and Commercial Perspectives*, 2nd ed. (Boston: Pearson, 1997)., Joseph B. Walther et al., "Attributes of Interactive Online Health Information Systems," *Journal of Medical Internet Research* 7, no. 3 (2005). https://doi.org/10.2196/jmir.7.3.e33., Kai Kuang, "The Role of Interactivity in New Media-Based Health Communication: Testing the Interaction among Interactivity, Threat, and Efficacy," in *Technology and Health*, eds. Jihyun Kim and Hayeon Song, Academic Press, 2020, 377–397, https://doi.org/10.1016/B978-0-12-816958-2.00017-4.

[82]Lisa P. Argyle et al., "Out of One, Many: Using Language Models to Simulate Human Samples," *Political Analysis* 31, no. 3 (2023): 337–351., Shriphani Palakodety, Ashiqur R. KhudaBukhsh, and J. Carbonell, *Mining Insights from Large-Scale Corpora Using Fine-Tuned Language Models*, Amsterdam: IOS Press, 2020, 1890–1897.

[83]Systems like GPT-Engineer (Osika, 2023) and BabyAGI (Nakajima, 2023) are examples of such systems from the software development domain.

[84]Argyle et al., "Out of One, Many: Using Language Models to Simulate Human Samples," 337–351., Shibani Santurkar et al., *Whose Opinions Do Language Models Reflect?*, arXiv, 2023, https://doi.org/10.48550/arXiv.2303.17548.

[85]Alec Radford et al., *Language Models Are Unsupervised Multitask Learners* (Open AI, 2019).

[86]Jacob Devlin et al., "BERT: Pre-Training of Deep Bidirectional Transformers for Language Understanding," in *Proceedings of the 2019 Conference of the North American Chapter of the Association for Computational Linguistics: Human Language Technologies* (Minneapolis, MN: Association for Computational Linguistics, 2019), 4171–4186, https://doi.org/10.18653/v1/N19-1423.

[87]Gati Aher, Rosa I. Arriaga, and Adam Tauman Kalai, *Using Large Language Models to Simulate Multiple Humans and Replicate Human Subject Studies*, arXiv, 2023, https://doi.org/10.48550/arXiv.2208.10264.

[88]Lisa P. Argyle et al., "Out of One, Many: Using Language Models to Simulate Human Samples," *Political Analysis* 31, no. 3 (2023): 337–351.

[89]AI Objectives Institute, *Introducing Talk To the City: Collective Deliberation at Scale*, 2023, https://www.talktothe.city/.

[90]Yonadav Shavit, *What Does It Take to Catch a Chinchilla? Verifying Rules On Large-Scale Neural Network Training Via Compute Monitoring*, arXiv, 2023, https://doi.org/10.48550/arXiv.2303.11341.

Existing SRM studies are split roughly evenly between fine-tuning and prompting. More recent approaches tend towards prompting, which tends to be used with larger models like GPT-3, though fine-tuning is also used more often with smaller models like BERT and GPT-2.

### 4.5. Inference Techniques

SRM practitioners employ a variety of open- and closed-ended inference techniques. In closed-ended inference, the model makes a prediction from a limited set of options chosen by the practitioner. Closed-ended inference often involves completing multiple-choice survey questions. For example, the authors in Chu et al. (2023) adapt survey questions to cloze-style tasks. Palakodety, KhudaBukhsh, and Carbonell (2020) and Argyle et al. (2023) frame vote prediction as a closed-ended inference task. Bisbee et al. (2023) use closed-ended generation to complete public opinion survey feeling thermometer responses. Kim and B. Lee (2023) use DCN architecture on top of LLM embeddings to predict General Social Survey responses, framed as a binary classification task.

In open-ended inference, model outputs are generally unconstrained. Argyle et al. (2023) also use open-ended text generation to collect partisan descriptors. Several studies (P. Feldman et al., 2022 P. G. Feldman, Pan, Shimei, and J. Foulds, 2023 H. Jiang et al., 2022) sample multiple open-ended texts from their tuned models, and apply sentiment analysis to the generated text, using aggregate sentiment as the final system output. We note that the use of this technique may be due to noisier performance of smaller GPT-2 models in comparison to larger models such as GPT-3. Open-ended generation is also used in M. A. Bakker et al. (2022) to generate consensus opinions. Dialog is a particular type of open-ended inference involving multiple conversational turns. The Talk to the City project (AI Objectives Institute, 2023) proposes to use SRMs for dialog. Bisbee et al. (2023) use the dialog interface to ChatGPT-3.5 for inference, but not in a multi-turn capacity. Simmons (2023) uses open-ended inference to generate moral rationalizations.

### 4.6. Evaluation Techniques

Several studies have proposed methods to evaluate SRM behavior. These can be broadly categorized into three types: direct comparison with ground truth, evaluation by human annotators, and qualitative (macroscopic) evaluation.

- **Ground truth comparison**: This defines an explicit function that compares ground truth response data with output from the SRM. This approach works well when representative ground truth data is available, and meaningful comparison metrics can be defined in the space of response variables. Humans may be involved in data collection but are not necessary to perform evaluation. This is sometimes referred to as *automatic evaluation* in the machine learning literature. For instance, Argyle et al. (2023) conducted a study where they used ANES Survey data along with demographic data to prompt GPT-3 to predict voting behavior. The predictions were then compared to actual partisan voting behavior, thus establishing a ground-truth comparison. H. Jiang et al. (2022) adopted a similar approach: they used the American National Election Study, which included items related to political figures and groups. Each item was labeled with a party based on favorability ratings among self-identified participants. The fine-tuned GPT-2 models were then prompted to generate open-ended text about each item, and the sentiment scores were calculated. These scores were used to determine GPT favorability, which was then compared to ground-truth favorability from the survey data.

[91]Rohan Taori et al., *Alpaca: A Strong, Replicable Instruction-Following Model* (Stanford Center for Research on Foundation Models, 2023), https://crfm.stanford.edu/2023/03/13/alpaca.html.

[92]Ben Wang and Aran Komatsuzaki, *GPT-J-6B: A 6 Billion Parameter Autoregressive Language Model*, 2021.

- **Human evaluation**: This applies a human-mediated evaluation to model outputs, either evaluating model outputs independent of any other data, or comparing the outputs to reference data. An example of this approach is the Turing-style experiments proposed in Argyle et al. (2023) , in which human raters are asked to judge whether a certain output is human or model-produced, and indistinguishability is treated as an indicator of success for the SRM. In the work of Chu et al. (2023) , various media such as online news articles, TV, and radio transcripts were used for fine-tuning their model. They evaluated their model's performance by measuring the correlation between model responses and human judgments in relation to the COVID-19 pandemic and consumer confidence. They used data from sources such as the Pew Research Center and the University of Michigan. M. A. Bakker et al. (2022) evaluate their consensus generation model by crowd worker comparison between their model outputs and other consensus generation baselines.

- **Macroscopic evaluation**: This compares SRM outputs and subpopulation characteristics qualitatively at an aggregate level rather than at the level of individual model outputs. Qualitative evaluation serves as a way to comprehend the model's outputs and subpopulation characteristics at an aggregate level, relying on expectations about aggregate behavior instead of ground truth data. This approach is useful when it is difficult to define meaningful quantitative metrics, or the phenomenon under study is complex enough that ground truth data collection and human evaluation are prohibitively expensive, or if the phenomenon can only be studied at an aggregate level, limiting the utility of statistical evaluation. This approach often involves visualization of SRM outputs.

As an example, Palakodety, KhudaBukhsh, and Carbonell (2020) conducted a qualitative evaluation to whether their SRM system responded to changes in political events over time. Simmons (2023) is another example, where the authors evaluate whether the moral biases of GPT and OPT LLMs agree with predictions about human behavior from Moral Foundations Theory (MFT), an influential theoretical framework in social psychology.[93] Moral content in LLM outputs conditioned on political identity are generated, and the author evaluates whether outputs agree directionally with results in humans.

### 4.6.1. Guiding Criteria

Existing work studying subpopulation representative models offers a number of criteria that guide their evaluations. Often, new terminology is proposed. This section synthesizes the evaluation criteria put forward in the surveyed work, collecting evaluation principles into five categories: *fidelity*, *necessity*, *robustness*, *sensitivity*, and *fairness*, with examples from existing work. We note that while the incipient nature of LLM technology undoubtedly presents new challenges for evaluation, many of the proposed criteria connect to familiar criteria for more traditional modeling approaches in the social sciences (as illustrated in Figure 3).

1. **Fidelity**: The fidelity criterion asks whether subpopulation representative models *successfully reproduce the target characteristics of the subpopulation they are designed to model.* We borrow this terminology from Argyle et al. (2023) whose definition of algorithmic fidelity is given in Section 3. For a subpopulation representation task, fidelity is perhaps the most salient criterion, as it essentially represents performance on the target task. Argyle et al. (2023) further subdivides fidelity into four criteria: social science Turing test, backward continuity, forward continuity, and pattern correspondence. We adapt these criteria from the social science domain to the task of subpopulation representative modeling, generalizing social science turing test to *indistinguishability*, generalizing forward and backward continuity to *consistency*, and using the notion of *pattern correspondence* essentially unchanged. We summarize existing SRM evaluation approaches belonging to the these three criteria (and a fourth—*diversity*) below.

---

[93]Jesse Graham et al., "Moral Foundations Theory: The Pragmatic Validity of Moral Pluralism," in , eds. Patricia Devine and Ashby Plant, Advances in Experimental Social Psychology, 2013, 55–130, https://doi.org/0.1016/B978-0-12-407236-7.00002-4.

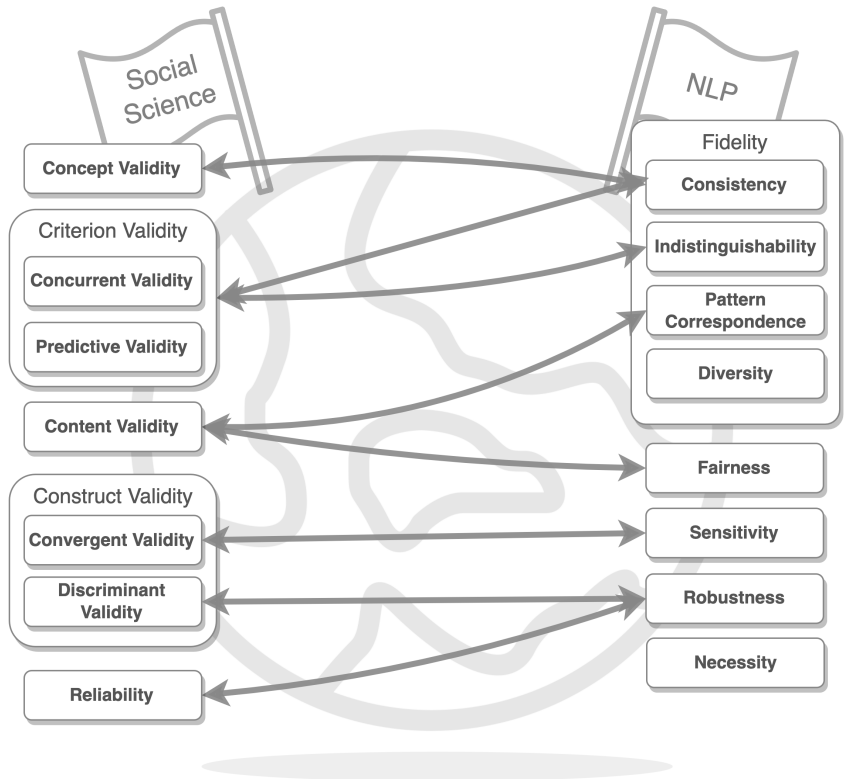

Figure 3: Guiding principles used to evaluate SRMs (right) and their connections to validity and reliability of social science measures (left).

Tests to measure fidelity are connected to criterion validity, and sometimes extend into construct validity depending on the abstract vs. concrete nature of the test. The most straight-forward setup to measure fidelity is testing predictive performance on out-of-sample data. This lines up with criterion validity, at least as "algorithmic modelers" see it.

(a) *Indistinguishability*. The indistinguishability criterion asks "whether the responses of subpopulation representative models can be distinguished from corresponding responses from the target subpopulation". This approach shares the philosophy of the Turing test[94]—the idea that computers will have successfully replicated some aspect of human behavior or intelligence when their responses are indistinguishable from human responses. Argyle et al. (2023) use this idea to evaluate the performance of GPT-3 prompted with partisan demographic information in open-ended descriptions of Republicans and Democrats, asking crowd workers to guess the source of human- and GPT-generated textual descriptions. Simmons (2023) uses this idea to evaluate LLM use of moral language, comparing between-humans and LLM-human differences in moral word use.

(b) *Consistency*. The consistency criterion asks whether subpopulation representative models behave consistent with the target subpopulation as well as the modeling task. This applies at a syntactic level, for example generating grammatically correct text continuations in response to a prompt, at a semantic level, for example choosing an answer from a list of multiple-choice answers rather than some other text. Consistency is generally a necessary condition for fidelity, but does not guarantee fidelity on its own.

Argyle et al. (2023) propose *forward continuity*, roughly meaning that model outputs should reflect the form, tone, and content of the context (the prompt). Aher, Arriaga, and Kalai (2023) use *validity* to refer to a similar concept. In their experiments using simulated behavior of humans undergoing psychological experiments, they define *valid* responses as responses that contain actions available to a participant in the simulated experiment.[95] These criteria resemble the familiar notion of *face validity*—whether a device subjectively appears to measure what it claims to measure.

Argyle et al. (2023) also propose *backward continuity*, roughly meaning that generated responses are consistent with the attitudes and sociodemographic content in the prompt, to such an extent that these conditioning variables can be inferred by humans observing only the model outputs. This seems closest to *construct validity* in the traditional taxonomy.

(c) *Pattern correspondence*. The pattern correspondence criterion, first proposed in Argyle et al. (2023) , asks "whether subpopulation representative models capture complex patterns between response variables in the target subpopulation".

(d) *Diversity*. The diversity criterion asks whether LLM outputs reflect the full range of subpopulation phenomena they intend to model. Bisbee et al. (2023) measure the variability in responses from multiple samples from ChatGPT conditioned with prompts drawn from ANES survey data, finding that the results overestimate the differences between groups, and underestimate the variance of opinions within a single demographic group.

2. **Necessity**: The necessity criterion asks "whether LLM-based methods are necessary for achieving satisfactory performance on the target subpopulation representative task". The rationale behind this criterion will be familiar to those with a background in machine learning, but perhaps unfamiliar for those with a background in statistical methods. Breiman (2001) describes the gap between these "Two Cultures."[96] One culture—the *data modelers*—makes *a priori* assumptions about the distributions underlying a data generation process, using a limited set of simple models that coincide with these assumptions. *Algorithmic modelers*, in contrast, make few assumptions about the data generation process, and select from a large space of model architectures and sizes that are often overparameterized by default. This leaves the algorithmic modeler to wonder whether a particular modeling approach (e.g., adding another layer to the neural network) is truly called for in a given modeling task. Increasing predic-tive performance on the training distribution is not, by itself, a reliable indicator, since the models quite regularly have enough capacity to overfit to the training distribution.[97] In addition to out-of-sample testing (i.e., cross-validation), machine learning practitioners commonly draw from a toolbox of techniques to gauge whether certain components of a model are truly necessary, and these same techniques have appeared in several SRM evaluations.[98]

---

[94]Alan M. Turing, "Computing Machinery and Intelligence," *Mind* 59 (1950): 433–460. https://doi.org/10.1093/mind/lix.236.433.

[95]For example, in a simulation of the Ultimatum Game from behavioral economics, valid completions in response to an offer of a certain amount of money must begin with accept or reject, the actions available to real human participants in the Ultimatum Game experiment.

One popular approach is to compare results from LLM-based SRMs to results obtained using simpler methods. For example, Chu et al. (2023) and Palakodety, KhudaBukhsh, and Carbonell (2020) compared results from BERT-based models to simpler $n$-gram language models, and H. Jiang et al. (2022) compare their results to a keyword retrieval baseline. Evaluating the necessity of a particular model size is another popular approach, since expense to develop and operate a model grows with the model size in number of parameters. Argyle et al. (2023) compared results from GPT-3 (175 billion parameters) to models as small as 125 million parameters, notably finding that competitive results are achieved by the much smaller GPT-Neo (2.7 billion parameters).[99] Simmons (2023) compares results from GPT and OPT models ranging from 350 million to 175 billion parameters.

Another popular approach is the ablation study. This approach studies system performance with various subcomponents, training steps, or techniques removed, deactivated, or replaced by simpler baselines, to evaluate the contribution of each component to overall system performance. Palakodety, KhudaBukhsh, and Carbonell (2020) and H. Jiang et al. (2022) compare model results with and without fine-tuning.

3. **Robustness**: The robustness criterion asks whether subpopulation representative models are robust to perturbations in the input data. Deep neural networks, including LLMs, are known to be vulnerable to *adversarial inputs* —inputs designed to be innocuous to human observers, while eliciting undesirable behavior from the models they are formulated against.[100] Additionally, neural networks including language models have been shown to be sensitive to spurious information in training data[101], as well as adversarial inputs.[102] While adversarial inputs are a concern for models that will be vulnerable to adversarial manipulation, the more general sense of robustness is important even for models that will only be used within a trusted organization.

Robustness directly impacts the viability of subpopulation representative models—for SRMs to be a viable approach, practitioners must be able to elicit model responses that maintain fidelity across regions of the input space, and that are not overly sensitive to details that would not be relevant to real individuals from the target subpopulation. The notion of robustness is connected to the notion of construct validity. If the SRM is robustly modeling the intended underlying phenomenon, then it should not respond to perturbations in dimensions with no relationship to that phenomenon. In particular, robustness connects to *divergent validity*—the ability of a measure to discriminate between related and unrelated concepts.

The importance of robustness as an evaluation criterion is evident in several studies. For example, Chu et al. (2023) emphasized the need for SRMs to be robust to re-phrasings that humans would not consider semantically different. They propose research questions to investigate robustness, such as whether media diet models are robust to paraphrases of survey question prompts, whether they have predictive power across media sources, and whether media diets remain predictive even when controlling for subpopulation demographics.

---

[96]See also Grimmer, M. E. Roberts, and Stewart (2021) .

[97]Trevor Hastie, Robert Tibshirani, and Jerome Friedman, *The Elements of Statistical Learning*, 2nd ed., Springer Series in Statistics (New York: Springer, 2009).

[98]Justin Grimmer, Margaret E. Roberts, and Brandon M. Stewart, "Machine Learning for Social Science: An Agnostic Approach," *Annual Review of Political Science* 24, no. 1 (2021): 1–25.

[99]Michiel A. Bakker et al., "Fine-Tuning Language Models to Find Agreement among Humans with Diverse Preferences," *Advances in Neural Information Processing Systems* 35 (2022): 38176–38189.

Similarly, Palakodety, KhudaBukhsh, and Carbonell (2020) highlighted the significance of robustness in SRMs and proposed robustness to rephrase as on operationalization of robustness. Additionally, they mention the need for robustness to catastrophic forgetting[103] (SRMs should maintain abilities after domain-specific fine-tuning) and negation (SRMs should "understand" negation words in their input).

4. **Sensitivity**: The sensitivity criterion asks whether subpopulation representative models display appropriate sensitivity[104] to their inputs. If SRMs are performant with a given amount of data, increasing the quality or quantity of the data should increase the performance of the model. An example is found in Chu et al. (2023) , where the authors hypothesize that their models should be more accurate for historical periods where more attention was paid to the media. This notion of sensitivity comfortably aligns with the traditional notion of construct validity—if an SRM is modeling what it sets out to model, then it should respond to perturbations in relevant variables. Between convergent and divergent (discriminant) validity, the sensitivity criterion appears conceptually related to convergent validity, which posits that a measure is convergently valid if it demonstrates a relationship to another measure in practice when this relationship is expected in theory.

5. **Fairness**: The concept of fairness has received much attention in machine learning literature. Several definitions have been proposed —Mehrabi et al. (2021) identify 10 distinct definitions of fairness, from three categories of individual fairness, group fairness, and subgroup fairness. Traditional definitions of fairness for machine learning systems tend to focus on the behavior of a single estimator, i.e., a single neural network.[105] These definitions require adaptation for the LLM setting, where prompting in combination with a single set of neural network weights can be used to solve widely varying tasks. Some work in this direction treats the combination of the model (the network weights) and the prompt together as the relevant system boundary for fairness evaluation.[106]

Kane (2010) comments on the intertwining nature of fairness and validity, highlighting that broad or narrow definitions for either term influence how the two relate to each other.[107] Kane argues that the concepts share substantial overlap, noting how a measure that systematically misrepresents some demographics on a construct is, to that extent, not valid for use with respect to those demographics. Fairness could be defined as equitable fidelity or criterion validity over all regions in the input space.

[100]Ian J. Goodfellow, Jonathon Shlens, and Christian Szegedy, "Explaining and Harnessing Adversarial Examples," in *3rd International Conference On Learning Representations, ICLR 2015, San Diego, CA, USA, May 7-9, 2015, Conference Track Proceedings*, eds. Yoshua Bengio and Yann LeCun (2015), https://doi.org/10.48550/arXiv.1412.6572.

[101]Suchin Gururangan et al., *Annotation Artifacts in Natural Language Inference Data*, arXiv, 2018, https://doi.org/10.48550/arXiv.1803.02324.

[102]Eric Wallace et al., "Universal Adversarial Triggers for Attacking and Analyzing NLP," in *Proceedings of the 2019 Conference On Empirical Methods In Natural Language Processing And the 9th International Joint Conference On Natural Language Processing (EMNLP-IJCNLP)* (Hong Kong: Association for Computational Linguistics, 2019), 2153–2162, https://doi.org/10.18653/v1/D19-1221.

[103]Robert M. French, "Catastrophic Forgetting in Connectionist Networks," *Trends in Cognitive Sciences* 3, no. 4 (1999): 128–135. https://doi.org/10.1016/S1364-6613(99)01294-2.

[104]Not to be confused with the mathematical notion of sensitivity, also known as the true positive rate.

Santurkar et al. (2023) evaluate whether models from AI21 labs and OpenAI attain equal fidelity when prompted to represent various subpopulations, Chu et al. (2023) evaluate whether "media diet model" fidelity varies by media diet, question topic, and question type, and H. Jiang et al. (2022) analyze errors of fine-tuned Republican and Democrat GPT-2 models across political figures and social groups from American National Election Study survey data.

### 4.7. MACHINE PSYCHOLOGY AND SIMULATING INDIVIDUALS WITH LLMS

The concepts of machine psychology and subpopulation representative modeling are closely related. Prior work has defined machine psychology as aiming to "elicit mechanisms of decision-making and reasoning in LLMs by treating them as participants in psychology experiments."[108] This definition naturally leans towards the ability of LLMs to simulate *individuals*. Like the authors in Hagendorff (2023) , Binz and Schulz (2023) used prompting as the elicitation method for representative behavior, but we imagine that representative behavior targeted as an individual may also be elicited by fine-tuning (for instance, on a corpus of one person's journal articles).[109] This raises the philosophical question of what LLMs are simulating.[110] Each individual is arguably defined by a lifetime of experience and an infinitude of characteristics that cannot fit in a fixed-size context window or be represented in a fixed number of training examples. If the prompt includes only a fixed number of demographic variables, is the LLM simulating an individual, the superposition of many individuals, or a subpopulation? What is the distinction between these three cases? We welcome work that clarifies these questions. Hagendorff (2023) covers the use cases for LLMs as tools for and objects of psychological study. We situate this representative ability in the broader context of decision-making use cases. Hagendorff also proposes methods to improve LLM performance in machine psychology (including avoiding training data contamination), but does not mention other vulnerabilities such as data poisoning.

### 4.8. IDENTIFIED RISKS

As with many impactful technologies, large language models pose various risks to individuals and societies that are affected by their use. Several authors have worked to taxonomize the risks posed by LLMs.[111] Subpopulation representative models using LLMs will inherit many of their risks. Proactive consideration of these risks may reduce the potential for harm. The studies we surveyed do draw attention to potential harms that could be introduced by SRM systems. In this section, we connect these risks to the taxonomy proposed in Weidinger et al. (2022) .[112]

H. Jiang et al. (2022) identified erasure of marginalized voices as a potential risk of the CommunityLM system, corresponding to the risk of

[105] Ninareh Mehrabi et al., "A Survey on Bias and Fairness in Machine Learning," *ACM Computing Surveys* 54, no. 6 (2021): 1–35. https://doi.org/10.1145/3457607.

[106] Huan Ma et al., *Fairness-Guided Few-Shot Prompting for Large Language Models*, arXiv, 2023, https://doi.org/10.48550/arXiv.2303.13217.

[107] Michael Kane, "Validity and Fairness," *Language Testing* 27, no. 2 (2010): 177–182. https://doi.org/10.1177/0265532209349467.

[108] Thilo Hagendorff, *Machine Psychology: Investigating Emergent Capabilities and Behavior in Large Language Models Using Psychological Methods*, arXiv, 2023, https://doi.org/10.48550/arXiv.2303.13988.

[109] Marcel Binz and Eric Schulz, "Using Cognitive Psychology to Understand GPT-3," *Proceedings of the National Academy of Sciences* 120, no. 6 (2023). https://doi.org/10.1073/pnas.2218523120.

[110] janus, *Simulators*, LessWrong, 2022, accessed October 19, 2023, https://www.lesswrong.com/posts/vJFdjigzmcXMhNTsx/simulators.

[111] Rishi Bommasani et al., *On the Opportunities and Risks of Foundation Models*, arXiv, 2022, https://doi.org/10.48550/arXiv.2108.07258., Renee Shelby et al., *Identifying Sociotechnical Harms of Algorithmic Systems: Scoping a Taxonomy for Harm Reduction*, arXiv, 2023, https://doi.org/10.48550/arXiv.2210.05791., Laura Weidinger et al., "Taxonomy of Risks Posed by Language Models," in *2022 ACM Conference On Fairness, Accountability, And Transparency* (New York: Association for Computing Machinery, 2022), 214–229, https://doi.org/10.1145/3531146.3533088.

[112] We use a single taxonomy for simplicity, and select Weidinger's taxonomy as it was qualitatively judged to offer the best fit to the surveyed studies.

"exclusionary norms" identified in Weidinger et al. (2022) Section 2.1.3. Likewise, M. A. Bakker et al. (2022) identified a risk for more authoritative or confident voices to dominate data collection, potentially exacerbating power imbalances. M. A. Bakker et al. (2022) , Chu et al. (2023) , P. G. Feldman, Pan, Shimei, and J. Foulds (2023) , and Talk to the City (AI Objectives Institute, 2023) identified model bias as a potential issue, corresponding to Sections 2.1.1 "Social stereotypes and unfair discrimination" and 2.1.4 "Lower performance for some languages and social groups" in Weidinger's taxonomy. (Kim and B. Lee 2023) also demonstrated higher performance in opinion prediction tasks for white, partisan, and high socioeconomic status demographics, highlighting reduced performance for marginalized demographics as a risk.

Potential for hallucination and *misinformation* ("Misinformation harms" in Section 2.3 of Weidinger et al., 2022) is identified in works by M. A. Bakker et al. (2022) , Argyle et al. (2023) , Chu et al. (2023) , P. Feldman et al. (2022) , and Talk to the City (AI Objectives Institute, 2023). The potential for LLM-based systems to produce *harmful text* is identified in works by Aher, Arriaga, and Kalai (2023) , Simmons (2023) , and P. Feldman et al. (2022) , corresponding to Weidinger's Section 2.1.2 on "Hate speech and offensive language". Performance issues related to *fidelity* are identified in Chu et al. (2023) , P. Feldman et al. (2022) , Santurkar et al. (2023) , and Talk to the City (AI Objectives Institute, 2023). These roughly correspond to 2.1.1 "Social stereotypes and unfair discrimination" and 2.1.4 "Lower performance for some languages and social groups" in Weidinger et al. (2022) . We can think of fidelity as leveraging an appropriate model bias after conditioning. Failure to elicit enough bias or bias in the appropriate direction, or elicitation of too much bias, as in Bisbee et al. (2023) , or bias in the wrong direction are two types of failures of algorithmic fidelity. Even if an SRM system does not suffer from risks associated with common failure modes of LLMs, this "success case" can still present a risk for SRM implementations, as the ability to model subpopulation characteristics presents opportunities for *misuse* including fraud and other types of harmful social manipulation. This aligns with Weidinger's section 2.4 "Malicious uses." A risk for misuse for persuasion was identified in M. A. Bakker et al. (2022) .

*Privacy* issues, such as the potential for LLMs to leak sensitive information including PII is identified in Aher, Arriaga, and Kalai (2023) , corresponding to Section 2.2.1 "Compromising privacy by leaking sensitive information" of Weidinger et al. (2022) . Kim and B. Lee (2023) also highlight the privacy and individual autonomy risks created by the ability to accurately predict opinions that survey respondents prefer not to disclose using imputation techniques, and the risk for misuse if organizations use this capability for decision-making.

Finally, some works identify *performance* issues that are unrelated to fidelity, including poor reasoning capabilities as a factor that may exacerbate other risks of SRM systems.[113]

## 5. Subpopulation Representative Models in Context

### 5.1. Predictive Performance as a Substituted Attribute

Opinion aggregation is challenging. Consider, for example, the results of the 2016 presidential election in the United States, where Republican candidate Donald Trump won against Democratic candidate Hillary Clinton, despite the majority of forecasting efforts which put the election in Clinton's favor.[114] It is s little wonder that opinion aggregation is turning to new sources of signal like social media and turning towards LLMs as a way to integrate this unstructured data for the task of opinion aggregation. If successful, this approach could overcome some of the inherent limitations in traditional approaches to polling and surveying. However, determining the right way to leverage LLMs for opinion aggregation is no small task. In part, this is because designing LLM-based applications is itself a challenging task. As Section 2 discussed, the LLM engineer is faced with a number of design choices related to model selection, elicitation method, and data, and this only scratches the surface of the complexity of the full endeavor. What can we expect from the practitioner faced with the doubly challenging task of leveraging LLMs for opinion aggregation?

---

[113]Philip Feldman et al., *Polling Latent Opinions: A Method for Computational Sociolinguistics Using Transformer Language Models,* arXiv, 2022, https://doi.org/10.48550/arXiv.2204.07483.

[114]Courtney Kennedy et al., "An Evaluation of the 2016 Election Polls in the United States," *Public Opinion Quarterly* 82, no. 1 (2018): 1–33.

In 2002, Kahneman and Frederick proposed *attribute substitution* as a fundamental mechanism underlying a host of familiar cognitive biases, such as availability and representativeness biases.[115] The basic idea of attribute substitution is that, when faced with the expensive evaluation of some *target attribute*, the mind's System 1 faculties instead evaluate a less demanding *heuristic attribute*, often without the individual's conscious awareness. While attribute substitution affects individuals, it is easy to notice that organizations and whole societies likewise substitute expensive evaluations for simpler proxy measures. A number of authors in management and organizational science have noted the tendency for organizations to optimize for quantitative "proxy measures" or "surrogates", and the pitfalls that come with this approach.[116]. Likewise, the mechanisms that have contributed to the scientific replication crisis can

be understood as forms of attribute substitution, at the scale of entire disciplines.[117]

We believe that this widespread tendency to measure and optimize for readily available heuristics has implications for the future of subpopulation representative models. When faced with the challenging evaluation of the *right* way to apply LLMs to the task of subpopulation representative modeling, it becomes tempting to swap in a less demanding attribute. When we consider attributes that are available for substitution, predictive performance is a likely candidate. The success of empirical machine learning over the last decade is evidence that individuals and labs myopically optimizing for improved scores on narrow benchmarks can cumulatively drive progress. However, like many heuristics, there are cases where focusing on predictive performance leads to outcomes that fail to satisfy our reflectively-endorsed values. Just because something can be predicted, does not mean it should (Figure 4).

Take, for example, algorithmic systems for recidivism prediction. The `COMPAS` algorithm, used by the U.S. court system to predict likelihood of repeated criminal offense, was the subject of a 2016 ProPublica investigation which found that "blacks are almost twice as likely as whites to be labeled a higher risk but not actually re-offend."[118] Subsequent studies have argued that ProPublica's analysis was flawed, and the algorithm remains the subject of controversy.[119] Even if such a system does achieve some threshold of predictive performance or fairness, *should* it be used? Should such a predictive task have been attempted in the first place? Developers of systems used in any high-stakes decision-making context will eventually be faced with these sorts of questions. As interest in subpopulation representative modeling grows, practitioners will face a natural temptation to focus on predictive performance now, and worry about challenging big-picture questions later. However, given the critical nature of opinion aggregation in democratic government, unanticipated outcomes could have significant consequences.

[115]Daniel Kahneman and Shane Frederick, "Representativeness Revisited: Attribute Substitution in Intuitive Judgment," in *Heuristics and Biases: The Psychology of Intuitive Judgment*, New York: Cambridge University Press, 2002, 49–81, https://doi.org/10.1017/CBO9780511808098.004.

[116]Robert D. Austin, *Measuring and Managing Performance in Organizations* (New York: Dorset House, 1996)., Jerry Z. Muller, *The Tyranny of Metrics* (Princeton: Princeton University Press, 2018).

[117]For example: citations and popularity as a proxy for validity, and the results of individual statistical tests as proxies for more thorough assessments of statistical conclusion validity.

[118]Jeff Mattu et al., *Machine Bias*, 2016, accessed October 20, 2023, https://www.propublica.org/article/machine-bias-risk-assessments-in-criminal-sentencing.

[119]We refer the reader to Brian Christian's (2020) account of the `COMPAS` controversy in Christian, *The Alignment Problem: Machine Learning and Human Values*, chapter 2.

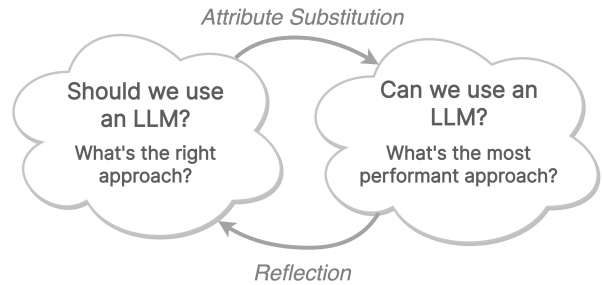

Figure 4: Predictive performance as a substituted attribute. This section warns practitioners in machine learning and political science against substituting the expensive evaluations (left) for their simpler counterparts (right).

What, then, is the alternative to quick-and-dirty reasoning based on substituted attributes? In the heuristics and biases literature, biases are overcome by deliberate practices involving reflection and structured thinking[120] we think that such an approach offers a worthwhile chance for machine learning practitioners and political scientists to avoid some of the less desirable consequences of hasty SRM development.

We next present two frameworks for structured thinking about subpopulation representative models. In Section 5.2, we take an *intra-system* approach, proposing a lens on subpopulation representative models through life cycle design, viewing the production of a model-based system as a life cycle from conception to eventual deprecation. Section 5.3 expands the scope further to an *inter-system* perspective, viewing SRMs through the lens of complex systems theory. We acknowledge that both of these analyses will be inherently incomplete. SRMs based on LLMs are still nascent, and we recognize that proposing a definitive analysis would be premature. Rather, we propose initial drafts, such that scholars from all relevant disciplines can contribute to a growing discussion of the requirements for responsible SRM development.

### 5.2. A Life Cycle Approach to Subpopulation Representative Models

Engineered systems tend to follow a common life cycle that starts with system design, followed by development and use, and eventually ends in deprecation. *Life cycle design* is an approach from systems engineering that designs systems with this full trajectory in mind. This approach offers benefits including clarity of communication around development processes, a roadmap to keep application development on track, reduction of risk from unplanned events during the development process, and transparency and auditability.[121] Naturally, various incarnations of the idea of life cycle design appear across the engineering disciplines. The *systems development life cycle* (SDLC) applies to software systems, and recent work has proposed a *machine learning life cycle* to structure the development of systems based on machine learning models and their accompanying data.[122] In the remainder of this section, we draw on existing SDLC and machine learning life cycle frameworks[123] to segment SRM development into three stages: *design*, *development*, and *operation*. These three stages are illustrated in Figure 5.

[120]Ozan Isler, Onurcan Yilmaz, and Burak Dogruyol, "Activating Reflective Thinking with Decision Justification and Debiasing Training," *Judgment and Decision Making* 15, no. 6 (2020): 926–938. https://doi.org/10.1017/S1930297500008147., Kathryn Ann Lambe et al., "Dual-Process Cognitive Interventions to Enhance Diagnostic Reasoning: A Systematic Review," *BMJ Quality & Safety* 25, no. 10 (2016): 808–820. https://doi.org/10.1136/bmjqs-2015-004417.

[121]David D. Walden, Garry J. Roedler, and Kevin Forsberg, "INCOSE Systems Engineering Handbook Version 4: Updating the Reference for Practitioners," *INCOSE International Symposium* 25, no. 1 (2015): 678–686.

### 5.2.1. Design

Subpopulation representative model implementation begins at the design stage, consisting of four steps: *task specification*, *system design*, *resource identification*, and *design evaluation*.

1. **Task (model) specification**: Task specification involves choosing a target subpopulation that the SRM will represent, and specifying a *predictive task*. The goal of an SRM is to estimate a set of *response variables* (opinions, attitudes, behaviors, or characteristics of the target subpopulation), from a set of *predictor variables* that will be used to condition the model. Specifying the predictive task involves selecting the predictor and response variables.[124]

2. **System design**: Once the practitioner has determined what subpopulation to model, and what the predictive task will be, they will need to determine how to approach the modeling task. The design stage proceeds with a design for the technical aspects of the SRM system: the choice of model architecture, selection of pre-trained or fine-tuned model checkpoints, and selection of a methodology to elicit subpopulation representative behavior. This elicitation method may be one or more fine-tuning methods, one or more prompting methods, or a combination thereof (see Sections 2.2 and 2.3 for an overview). Examples from existing SRM implementations include fine-tuning on subpopulation data as in H. Jiang et al. (2022) , RLHF with human annotators from the target subpopulation as in M. A. Bakker et al. (2022) , or conditioning the model with prompts derived from subpopulation data as in Argyle et al. (2023) .[125]

3. **Resource identification (data design)**: This stage includes identification of pre-existing data or plans to generate bespoke data that will be used to elicit subpopulation representative behavior. Any SRM implementation that leverages data for elicitation will involve some process to obtain this data, either from pre-existing sources or via a purpose-specific data generation process. Data for current SRMs comes from a variety of sources, including social media (Palakodety, KhudaBukhsh, and Carbonell, 2020; H. Jiang et al., 2022; P. G. Feldman, Pan, Shimei, and J. Foulds, 2023; P. Feldman et al., 2022), news media collections

[122]Rob Ashmore, Radu Calinescu, and Colin Paterson, "Assuring the Machine Learning Lifecycle: Desiderata, Methods, and Challenges," *ACM Computing Surveys* 54, no. 5 (2022): 1–39. https://doi.org/10.1145/3453444., Ben Hutchinson et al., "Towards Accountability for Machine Learning Datasets: Practices from Software Engineering and Infrastructure," in *Proceedings of the 2021 ACM Conference On Fairness, Accountability, And Transparency* (New York: Association for Computing Machinery, 2021), 560–575, https://doi.org/10.1145/3442188.3445918.

[123]Bo Li et al., "Trustworthy AI: From Principles To Practices," *ACM Computing Surveys* 55, no. 9 (2023): 1–46. https://doi.org/10.1145/3555803., Samuli Laato et al., "AI Governance in the System Development Life Cycle: Insights on Responsible Machine Learning Engineering," in *Proceedings of the 1st International Conference On AI Engineering: Software Engineering For AI* (New York: Association for Computing Machinery, 2022), 113–123, https://doi.org/10.1145/3522664.3528598., Saleema Amershi et al., "Software Engineering for Machine Learning: A Case Study," in *2019 IEEE/ACM 41st International Conference On Software Engineering: Software Engineering In Practice (ICSE-SEIP)* (2019), 291–300, https://doi.org/10.1109/ICSE-SEIP.2019.00042., Ashmore, Calinescu, and Paterson, "Assuring the Machine Learning Lifecycle: Desiderata, Methods, and Challenges," 1–39.

[124]In most social science fields, this step is commonly referred to as "model specification" or "model selection."

[125]Prompt tuning methods may also be used, although we are not aware of any existing SRMs that use prompt tuning.

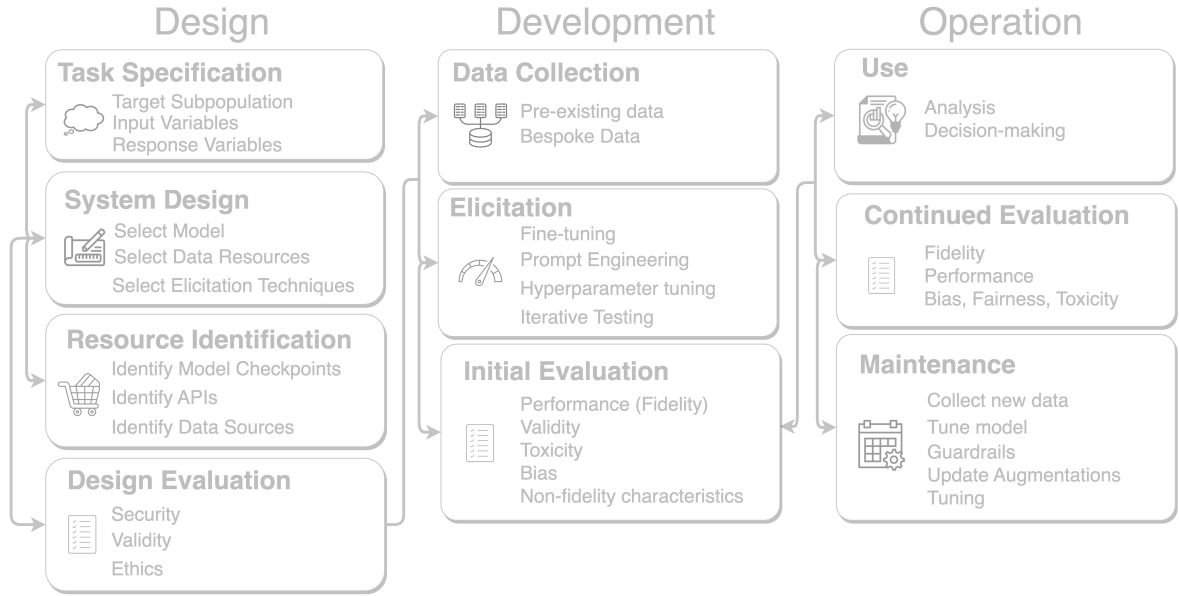

Figure 5: Subpopulation representative model (SRM) life cycle. SRM implementation begins with model design, including the choice of model architecture, selection of pretrained or fine-tuned model checkpoints, and sourcing of data that will be used to elicit representative behavior via fine-tuning or prompting. Elicitation consists of obtaining representative behavior from the model, using any combination of elicitation techniques including self-supervised, supervised, or reward model-based fine-tuning, prompting, and prompt tuning. Evaluation consists of measuring the functionality of the resulting SRM, including its fidelity to the target subpopulation, and other behavior including performance characteristics and potential to generate harmful output.

(Chu et al., 2023), and survey data (Chu et al., 2023; Argyle et al., 2023).

4. **Design evaluation**: System design is often followed by an initial evaluation. While the use of such a stage varies by the type of engineering project, and is not necessarily ubiquitous in life cycle design, we include it here as it presents an opportunity for feedback from groups other than those directly involved in the SRM implementation, prior to any investment in system development, but after there is a more detailed understanding of what the system will look like. The design evaluation stage is an opportunity to assess the proposed system for a number of characteristics:

- Does it use protected attributes for prediction or decision-making?

- What will be required to collect the requisite data?
- What existing systems will it replace, who will be affected?
- How much will it cost?
- Who should be consulted about the system development? This includes not only traditional stakeholders (consumers for traditional polling), but also parties whose data may be used to train the system.

System design evaluation before implementation should be familiar to practitioners in the social sciences, who often first have studies reviewed by an institutional review board prior.

### 5.2.2. Development

The development phase of the SRM life cycle consists of collecting data as specified in the system design, tuning the model to elicit subpopulation representative behavior and other desired behaviors, and evaluating the tuned model for its desired performance characteristics.

1. **Data collection**: Data collection consists of obtaining the data that will be used for fine-tuning, prompting, or other inference techniques, as well as data that will be used for evaluation of model performance (fidelity) and other important model behaviors (e.g., reasoning capabilities, bias, toxicity, etc.). This may involve obtaining datasets identified during the design stage, or the construction of datasets i.e., via polling, crowdsourcing, as planned during the design stage. To illustrate the diversity in implementations at this stage, Palakodety, KhudaBukhsh, and Carbonell (2020) and H. Jiang et al. (2022) obtain their data from social media, while M. A. Bakker et al. (2022) obtain data from a crowdsourced human feedback collection process.

2. **Elicitation (training/tuning)**: The tuning stage consists of applying the elicitation technique selected during the design stage to elicit subpopulation representative behavior from the model. Often at this stage, a variety of models of varying configurations will be tuned and evaluated. This process is often referred to as *hyperparameter optimization*.

3. **Initial evaluation**: Prior to deployment for end use, it is common for machine learning models to undergo a battery of tests that measure various model characteristics, including generalization performance.[126] Similar evaluations are likely in the SRM setting. The nature of the subpopulation representative modeling task introduces the need for unique evaluation criteria not used in other tasks. See Section 4.6 for an overview of techniques currently employed to evaluate SRMs, and Section 4.6.1 for connections to familiar concepts of reliability and validity. In the typical machine learning life cycle, the initial evaluation ends positively (where the model comes close enough to satisfying the design criteria and is fit for deployment) or negatively, often resulting in further exploration of the system design space or additional data collection.

### 5.2.3. OPERATION

The use or deployment stage of the SRM life cycle consists of using the developed model for the target task, regularly evaluating model performance, and performing necessary maintenance on the model and surrounding infrastructure. Depending on the use case, the model may be made available to a closed group of practitioners, or to the subpopulation as a whole. Interactions with the model range from simple regression or classification prediction familiar to social scientists, to interacting with an SRM through a dialogue interface similar to ChatGPT, to interacting with an agentic SRM through interfaces similar to AutoGPT.

1. **Use for the target task**: Currently, much of the work on SRMs is academic work focused on evaluating their potential for future use. However, there are already some projects who propose to using SRMs for decision making and community management at national and city levels (e.g., projects such as ION and Talk to the City).[127] See Section 4.3 for a review of tasks for which current SRMs have been used or proposed.

   We expect that SRMs will see use in both closed and open-access formats, and at varying levels of task sophistication, from more conventional regression and classification models not dissimilar to existing polling techniques, to LLM-based agents that take complex representative action on the behalf of subpopulations. At this stage, depending on the use case, the SRM will have impacts on the public, including the subpopulation it is designed to represent.

---

[126]Saleema Amershi et al., "Software Engineering for Machine Learning: A Case Study," in *2019 IEEE/ACM 41st International Conference On Software Engineering: Software Engineering In Practice (ICSE-SEIP)* (2019), 291–300, https://doi.org/10.1109/ICSE-SEIP.2019.00042.

2. **Continued evaluation**: Supporting use of the model for the intended task is often not the only obligation for the practitioner after a model has been developed. Machine learning model deployment typically subjects the model to various distribution shifts, where the input data the model and practitioner see at deployment time differs in some meaningful way from the input data during training and development. The most basic case is generalization to the test set. Often, the value of a predictive model is that it can replace a more expensive process of observing variables in the real world, by generalizing to input data that was unseen during development.

3. **Maintenance/iteration**: In practice it is common for deployed machine learning systems to be continuously maintained as requirements change, the input data distribution changes, or use of the model changes. Specific tasks at this stage can include retraining or continued training to arrive at new model parameters, updating or re-engineering prompts, updating data sources that are used to provide context for text-generation, updating guardrails on the model output, or downsizing the model via techniques like distillation to reduce inference costs. We can anticipate sampling more subpopulation data to update the model as current events change and opinions evolve. In the case that SRMs are used for decision-making, this means that the SRM and the opinions drawn from the lived experience of the target subpopulation may be both downstream and upstream of each other, forming a feedback loop. Feedback loops are one of the hallmarks of complex systems, and it is expected that these inter-system effects will be a key area of consideration for LLM-based subpopulation representative systems. We explore the role of SRMs as interactants in a larger complex system in the next section.

### 5.3. A Complex Systems Approach to Subpopulation Representative Models

Once a model-based system is developed and deployed, it often becomes an interactant in several surrounding complex systems. In the case of subpopulation representative modeling, these surrounding complex systems include opinion aggregation, governance, and democracy itself.

A *complex system* is a system of many interacting components, where the behavior of the whole system is more than the sum of its parts.[128] These systems tend to exhibit several key characteristics. Interactions between system components are generally nonlinear—changing one aspect of the system behavior can have a proportional effect, outsize effect, or no effect at all. Complex systems typically contain positive and negative feedback loops. Second-order (and higher-order) effects cannot be ignored, and often dominate the system dynamics. Although complex systems—unlike chaotic systems—exhibit long-term behavioral stability, the effect of unprecedented interventions on the system are nonetheless notoriously difficult to predict.

The study of complex systems spans a number of scientific disciplines. Perhaps most relevant to the present scope is the study of *sociotechnical systems*—complex systems consisting of both human and technological factors.[129] A full sociotechnical systems analysis of subpopulation representative modeling is beyond the scope of this review, although critically important. To motivate interdisciplinary study on this topic, we provide brief examples of previous cases of technological impacts on opinion aggregation and political systems. In particular, we emphasize the unantici-

[127] *ION - Primul Consilier Cu Inteligență Artificială Al Guvernului*, 2023, https://ion.gov.ro/., AI Objectives Institute,

*Introducing Talk To the City: Collective Deliberation at Scale*, 2023, https://www.talktothe.city/.

[128] James Ladyman, James Lambert, and Karoline Wiesner, "What Is a Complex System?," *European Journal for Philosophy of Science* 3 (2013): 33–67.

[129] Donald Martin et al., *Extending the Machine Learning Abstraction Boundary: A Complex Systems Approach to Incorporate Societal Context*, arXiv, 2020, https://doi.org/10.48550/arXiv.2006.09663.

pated second-order consequences that occur when new technology is infused into an existing part of the macro-level system of governance and politics. Our hope is that these examples will encourage practitioners to consider what we stand to gain by taking an integrative, precautionary stance on SRM development—or what we stand to lose by rushing ahead.

### 5.3.1. Opinion Aggregation as a Complex System

A first example comes from the interaction between polling technology and declining response rates. Groves (2011) divides the history of survey research into three eras: 1930s-1960s response rates were high, and surveys were conducted primarily in-person or by mail. As telephone surveying and computer-aided analysis became more prevalent, 1960-1990 saw increased ease in data collection and analysis and refinement of techniques from the previous era. Yet this era also saw the beginning of declining response rates. This decline steepened in the years following 1990 with the introduction of the Internet into polling and surveying. Leeper (2019) conjectures that, like the fishing industry, polling relies on a shared resource pool of responses from survey participants. As new technology increased contactability, public opinion was solicited more frequently, and people became less likely to respond to any individual request. Despite new technologies being incredibly useful to *increase* reachability and analytical sophistication, opinion polling also suffered from a *decrease* in response rates through the eras.

We present this case study because it exemplifies a pattern common to interventions in complex systems (as illustrated in Figures 6–8). First, a new tool or technology is introduced. Second, the new technology provides useful capabilities that appear to be aligned with the goals of practitioners in the field. Third, the new capabilities lead to desirable first-order consequences—improved reachability, for example. Fourth, the intervention yields unintended second-order consequences that are *misaligned* with the goals of practitioners in the system.

Figure 6: Polling technologies and response rates as an example of second-order effects of new technologies in opinion aggregation.

New technology gave pollsters increased capability to solicit opinions from the public and analyze the results, facilitating the 1930-1960 "era of expansion" noted by Graves. This first-order effect of increased data collection due to ease of contact was followed by the second-order effect that the shared resource of poll responses became overutilized, and response rates plummeted. While new technology was not the whole story, there is a case to be made that the same technologies that made opinion aggregation more effective over a short horizon paradoxically made it more difficult over a longer time horizon.

Figure 7: Social media and democracy as an example of second-order effects of new technologies in opinion aggregation.

Social media provides another instructive example of this pattern. Many will recall Western narratives about the role of social media in the "Arab Spring", a series of political uprisings in the Middle Eastern states, of Tunisia, Egypt, Yemen, Libya, Syria, and Bahrain, expressing popular frustration about authoritarian government, corruption, and economic stagnation. Social media was highlighted in the West as instrumental to this wave of political action.[130] These narratives appear in stark contrast with attitudes about its recent role in American political discourse—once hailed as a vehicle for democracy, social media can now just as easily be seen as a destabilizing force and a threat vector for adversarial influence.[131]

If we view subpopulation representative modeling through this lens, the stage appears set. The new technology of LLMs clearly offers new capabilities for opinion aggregation, notably the ability to process much larger quantities of unstructured text data and to interrogate them via open-ended language interface. The papers we surveyed highlight the potential first-order benefits of these new capabilities, and there is already budding interest to deploy such systems. What can we expect in the way of second-order consequences? While the specifics may be hard to predict, we can speculate and ask some reflective questions in an attempt to anticipate likely outcomes.

First, we can consider unanticipated consequences from an intra-system perspective; those conse-quences arising from the SRM itself. Model biases from the pretrained model checkpoint may propagate into the final SRM.[132] Our evaluations during model development might not include more than aggregate measures of fidelity, so we might miss particularly egregious bias for one subgroup of the overall demographic we intend to model. The model might hallucinate an opinion not representative of the underlying subpopulation.

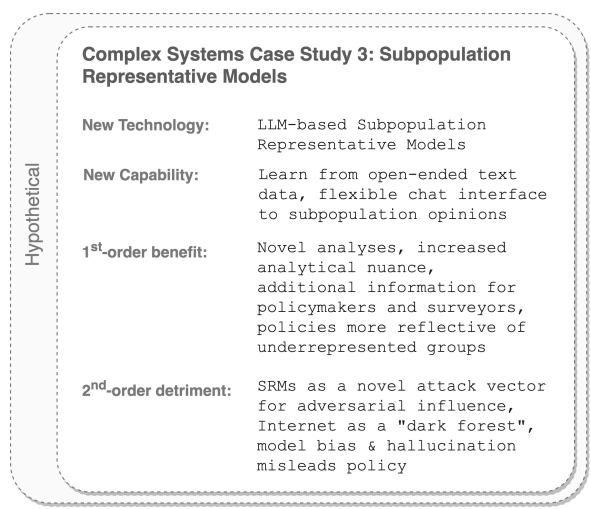

Figure 8: Hypothetical, unanticipated second-order effects of new technologies in opinion aggregation.

What happens to responses when citizens are aware that their activity is being fed into an LLM? Savvy internet users are already adding text snippets to websites that are invisible to human viewers but visible to retrieval-augmented LLMs such as Bing Chat. While this phenomenon is currently in its "toy" phase, more serious uses are not hard to imagine. What happens when adversaries are aware of LLMs as a factor in political action? The vulnerability of open-source models and online training loops has already been demonstrated, through backdoor attacks on open-source models[133], as well as a recent study showing that a small number of adversarial samples introduced during RLHF data collection can significantly degrade model performance for target keywords.[134] Forgetting about foreign adversaries, what happens when domestic politicians see LLMs as part of the information infrastructure? How hard would it be to rally one's supporters on Twitter or Facebook to distort the public opinion signal being fed into LLMs?

[130]Philip N. Howard and Muzammil M. Hussain, *Democracy's Fourth Wave? Digital Media and the Arab Spring* (New York: Oxford University Press, 2013).

[131]In 2019, the US Department of Justice released findings from an investigation into the 2016 election, which found a "sweeping and systemic influence campaign" by Russian state actors, including organic activity and ad spending on Facebook and other social media platforms (R. S. Mueller 2019). Even without adversarial involvement, several authors have highlighted the potential detriment of social media on American political discourse, noting the ability of social media to create echo chambers (Cinelli et al., 2020; Garimella et al., 2018), and the potential for ad-based revenue models combined with recommendation algorithms to promote inflammatory content (Munn, 2020; Merrill and Oremus, 2021; Whittaker et al., 2021).

[132]Shangbin Feng et al., "From Pretraining Data To Language Models To Downstream Tasks: Tracking The Trails Of Political Biases Leading To Unfair NLP Models," in *Proceedings of the 61st Annual Meeting Of the Association For Computational Linguistics (Volume 1: Long Papers)* (Toronto, Canada: Association for Computational Linguistics, 2023), 11737–11762, https://aclanthology.org/2023.acl-long.656.

## 5.4. Ethical Considerations

This work reviews existing implementations of SRMs and many of the existing implementations identify ethical concerns related to SRMs, as discussed in Section 4.7. We believe that integration of SRMs into social decision-making infrastructure presents additional risks related to epistemic security.[135] LLMs have been shown to be vulnerable to adversarial influence[136], a risk not discussed in any of the SRM implementations reviewed in this paper. For example, a recent study showed that with as few as 100 adversarial examples introduced during instruction-tuning, adversaries can severely degrade LLM performance for tasks involving a trigger phrase, e.g., "Joe Biden."[137] We believe SRM practitioners should take these additional risks into consideration. Like most technologies, SRMs have dual use potential—we recognize promising applications for SRMs in providing targeted representation of currently underrepresented groups through low cost and potential for targeted analysis. However, this depends on careful implementation of the SRM systen.

## 5.5. Takeaways

It is important to stress that the point in asking these questions is not to argue about whether subpopulation representative models are inherently good or bad, nor is it to argue unilaterally for or against their use. The point of this section is that: it should be taken as the null hypothesis that introducing SRMs into the complex system of democracy will have unanticipated consequences.

Consequently, we believe that the study of subpopulation representative models should focus not only on revealing their intra-system limitations (e.g., bias[138] and extremism[139]), but should also focus on how to fix these limitations to make them fit for their first-order uses. Additional effort should be applied to understand how to anticipate their second-order consequences and, ideally, how to avoid the least desirable of these. This requires machine learning experts, political scientists, ethicists, and experts from a diverse array of other domains to work across disciplinary boundaries.

Some readers may object that current SRMs are only prototypes, intended as objects for scientific study, not as production opinion aggregation systems. Indeed, current SRMs are mostly scientific prototypes, and deficient in some aspects that would be necessary to use them as robust instruments of political analysis. Does this mean that it is too early to consider the effects of subpopulation representative models at scale? Two proposals (Talk to the City and ION) already demonstrate interest for SRMs. Additionally, we would like to echo the words of John Gall, *The Systems Bible: The Beginner's Guide to Systems Large and Small*: "A temporary patch will very likely be permanent." Software developers sometimes informally refer to this phenomenon as *proto-duction*—a prototype that ends up with a lasting role in production despite being designed for temporary use. Even studies that intend to probe the capacities of a system for the sake of scientific interest can end up reimplemented as production systems if there is demand for said capabilities, even when the system is far from perfect. Only time will tell whether this widespread adoption will actually happen. In any case, we argue that there is enough of a possibility that it is worthwhile for SRM practitioners to anticipate the role of SRMs as part of the larger complex system of democracy.

[133]Daniel Huynh and Jade Hardouin, *PoisonGPT: How We Hid a Lobotomized LLM on Hugging Face to Spread Fake News*, July 9, 2023, https://blog.mithrilsecurity.io/poisongpt-how-we-hid-a-lobotomized-llm-on-hugging-face-to-spread-fake-news/.

[134]Alexander Wan et al., *Poisoning Language Models during Instruction Tuning*, arXiv, 2023, https://doi.org/10.48550/arXiv.2305.00944.

[135]Elizabeth Seger et al., *Tackling Threats to Informed Decision-Making in Democratic Societies: Promoting Epistemic Security in a Technologically-Advanced World* (London: The Alan Turing Institute, 2020).

[136]El-Mahdi El-Mhamdi et al., *SoK: On The Impossible Security Of Very Large Foundation Models*, arXiv, 2022, https://doi.org/10.48550/arXiv.2209.15259.

[137]Wan et al., *Poisoning Language Models during Instruction Tuning*.

[138]Shibani Santurkar et al., *Whose Opinions Do Language Models Reflect?*, arXiv, 2023, https://doi.org/10.48550/arXiv.2303.17548.

[139]James Bisbee et al., *Artificially Precise Extremism: How Internet-Trained LLMs Exaggerate Our Differences*, SocArXiv, 2023, https://doi.org/10.31235/osf.io/5ecfa.

## 6. Future Work

There is growing interest in LLMs as subpopulation representative models. While LLMs do show promise for this use case, they might not yet offer a comprehensive solution for the needs of political analysis. Additionally, the inherent limitations of LLMs may present SRM practitioners with new challenges. In this section, we discuss the future of subpopulation analysis tasks with LLMs. We cover additional types of political analyses and how LLMs might be applied to them (Section 6.1) and draw attention to the importance of benchmarking and evaluation (Section 6.2). We also comment on model interpretability (Section 6.3), consider whether LLMs might solve the sparse cell problem (Section 6.4), and conjecture about how sampling methodology might be developed for LLMs (Section 6.5).

### 6.1. Applying LLMs to Other Types of Political Analysis

Most of the surveyed works applied LLMs to tasks where the goal is aggregating subpopulation sentiment or simulating subpopulation behavior. In most cases, subpopulation identifiers were determined a priori. While this approach (analogous to polling and survey research) is popular in political science, it is not the only kind of analysis that political scientists might be interested in. For example, subgroup detection is another popular analytical technique, where the goal is to discover subgroups in some data from the larger population.

Traditionally, social scientists have used unsupervised learning techniques to identify substantively interesting (but unknown) subpopulations in the mass public.[140] For instance, J. L. Hill and Kriesi (2001) use mixture modeling to classify survey respondents into three groups (opinion holders, vacillating changers, and durable changers) based on their response patterns. Likewise, Fowler et al. (2023) use a mixture item response theory (IRT) model to estimate the proportion of voters whose political attitudes are: (1) structured unidimensionally (left-right); (2) structured, but multidimensionally; and (3) poorly structured in either ideological space. Cluster analysis and latent class analysis are especially popular specific methods for identifying latent or "hidden" subgroups in public opinion data. For example, these subgroups might differ in their types of political participation[141], attitude structures[142], political tolerance[143], ideological beliefs[144], or climate change attitudes.[145]

[140]Philip D. Waggoner, *Modern Dimension Reduction*, Elements in Quantitative and Computational Methods for the Social Sciences (New York: Cambridge University Press, 2021), https://doi.org/10.1017/9781108981767.

[141]R. Michael Alvarez, Ines Levin, and Lucas Núñez, "The Four Faces of Political Participation in Argentina: Using Latent Class Analysis to Study Political Behavior," *Journal of Politics* 79, no. 4 (2017): 1386–1402.

[142]John A. Fleishman, "Types of Political Attitude Structure: Results of a Cluster Analysis," *Public Opinion Quarterly* 50, no. 3 (1986): 371–386.

[143]Allan L. McCutcheon, "A Latent Class Analysis of Tolerance for Nonconformity in the American Public," *Public Opinion Quarterly* 49, no. 4 (1985): 474–488.

[144]Stanley Feldman and Christopher Johnston, "Understanding the Determinants of Political Ideology: Implications of Structural Complexity," *Political Psychology* 35, no. 3 (2014): 337–358., Justin H Gross and Daniel Manrique-Vallier, "A Mixed Membership Approach to Political Ideology," in *Handbook of Mixed Membership Models and Their Applications*, eds. Edoardo M. Airoldi et al., Boca Raton, FL: CRC Press, 2015, 119–140.

[145]Sam Crawley, Hilde Coffé, and Ralph Chapman, "Public Opinion on Climate Change: Belief and Concern, Issue Salience and Support for Government Action," *The British Journal of Politics and International Relations* 22, no. 1 (2020): 102–121.

LLMs may be able to offer new capabilities for subgroup detection. This direction is explored in Feldman's (2023) Keyword Explorer Suite software[146], where the user iteratively develops a keyword list by viewing social media search results based on the list, using the final corpus of social media posts to train an SRM. In the field of machine learning, it is already common practice to perform unsupervised learning on LLM embeddings.[147] This approach allows for the representation of unstructured data in a shared feature space, so that it may be used as input to traditional machine learning algorithms. Recent work has used the in-context learning ability of LLMs and the recent increases in context window size to perform unsupervised group detection entirely through prompting.[148] Rather than embedding each data point separately, all data points are presented to the LLM, along with a goal for the unsupervised analysis. This approach offers increased ability to focus the unsupervised analysis on particular analytical goals via prompting. However, the approach is limited by the context length of the LLM, may be less interpretable than some traditional group discovery approaches, and is more computationally expensive than traditional forms of clustering. Moreover, a fundamental is-sue here is that subgroups (whether latent are observed) are often interesting precisely because their opinions are anomalous. This will be problematic in situations where LLMs artificially reduce the variance of simulated responses (e.g., Bisbee et al., 2023).

In addition to—or in conjunction with—detecting substantively interesting subgroups, LLMs may also be able to identify emerging controversies or cleavages in public opinion. That is, LLMs may help to detect issues that are in the early stages of the "issue evolution" process described by Carmines and Stimson (1989) but for which survey response data is not widely available.[149] It is noteworthy that LLMs, while excelling in imputation and retrodiction tasks with public opinion survey data, also perform modestly for zero-shot prediction tasks.[150] In a zero-shot prediction task, the algorithm makes a prediction for a novel survey question—that is, one for which no responses are available in the training data. While Kim and B. Lee (2023) emphasize that LLM performance is limited when comparing the zero-shot predictions with true responses in the 1972-2021 General Social Survey data ($r = 0.67$ and AUC-ROC = 0.729), this nonetheless provides a remarkable tool for researchers piloting new survey questions and generating hypotheses—one for which there is currently no readily available alternative. LLMs also provide a novel way to address biases in public opinion surveys that emanate from agenda effects, in which our measurement of quantities like ideological polarization is but a reflection of the kinds of issues we include.

[146]Philip G. Feldman, Shimei Pan, and James Foulds, "The Keyword Explorer Suite: A Toolkit for Understanding Online Populations," in *Companion Proceedings of the 28th International Conference on Intelligent User Interfaces* (New York: Association for Computing Machinery, 2023), 21–24, https://doi.org/10.1145/3581754.3584122.

[147]Roee Aharoni and Yoav Goldberg, "Unsupervised Domain Clusters in Pretrained Language Models," in *Proceedings of the 58th Annual Meeting of the Association for Computational Linguistics* (Online: Association for Computational Linguistics, 2020), 7747–7763, https://doi.org/10.18653/v1/2020.acl-main.692., Zihan Zhang et al., "Is Neural Topic Modelling Better Than Clustering? An Empirical Study on Clustering with Contextual Embeddings for Topics," in *Proceedings of the 2022 Conference of the North American Chapter of the Association for Computational Linguistics: Human Language Technologies* (Seattle, WA: Association for Computational Linguistics, 2022), 3886–3893, https://doi.org/10.18653/v1/2022.naacl-main.285.

[148]Ruiqi Zhong et al., *Goal Driven Discovery of Distributional Differences Via Language Descriptions*, arXiv, 2023, https://doi.org/10.48550/arXiv.2302.14233.

[149]Indeed, although it is primarily an elite-driven account of partisan realignment, Carmines and Stimson also note that issue evolutions often arise from the policy concerns of issue publics: subpopulations for whom certain policy issues are especially relevant and salient (see also Krosnick, 1990).

[150]Junsol Kim and Byungkyu Lee, *AI-Augmented Surveys: Leveraging Large Language Models for Opinion Prediction in Nationally Representative Surveys*, arXiv, 2023, https://doi.org/10.48550/arXiv.2305.09620.

### 6.2. Evaluation and Benchmarking

Development of benchmarks and evaluation methodologies for SRMs will be a critical step in their responsible development and use. We expect that development of rigorous benchmarks for subpopulation representative modeling may prove challenging, especially for higher-complexity SRM tasks such as dialogue and agentic behavior (Figure 2). Open-ended tasks cannot be easily evaluated with traditional closed-ended metrics like accuracy. In the field of natural language processing, the difficulty to evaluate open-ended output has led towards using models to evaluate other models. This line of work includes examples like `BERTScore`[151] for textual similarity, and includes more recent work where LLMs are used to generate benchmark datasets.[152]

One important theme in evaluating LLMs is the potential for *data memorization*. In machine learning, models are typically tested on data that was not used to produce the model to assess *generalization*. Benchmark datasets in natural language processing are often split into train and test portions, where practitioners agree to use the train portion to develop their models, and the test portion to evaluate their generalization. The accidental inclusion of test data during training is often referred to as *data leakage*, and can inflate performance metrics, causing the evaluator to mistake *generalization* for performance on already-seen examples, or *memorization*.

With pre-training corpora growing to comprise larger and larger swaths of the internet, it is difficult to know whether any of the data used for evaluation was present in the pretraining corpus, (i.e., whether data leakage has occurred). The severity of this problem has led some practitioners to call for test portions of NLP benchmark datasets not to be uploaded in plain-text formats.[153] This issue touches all aspects of LM evaluation, including the evaluation of SRMs. Without access to the training data for GPT-3, for example, it is difficult to know whether survey responses to the American National Election Study (like those used for validation in Argyle et al. (2023) , or results from landmark sociological analyses (like those used for validation in Aher, Arriaga, and Kalai (2023) were present in the pre-training data. In other words, it is difficult to know whether strong performance of a model like GPT-3 on these benchmarks indicates *generalization* or *memorization*.

### 6.3. Model Interpretability

Grimmer et al. (2021) observed that if social scientists have been slow to adopt predictive modeling techniques, it is for good reason—sophisticated "black-box" modeling techniques can be notoriously difficult to interpret. The demand for interpretability in the social sciences may make LLMs attractive as an engine for subpopulation modeling. By operating directly on natural language, LLMs appear to remove much of the complexity of previous paradigms of predictive modeling. A practitioner prompting LLMs need not perform feature engineering or tune dozens of hyperparameters. Demographics can be added at will by extending a natural language description. With techniques like chain-of-thought prompting[154] or tree-of-thought prompting[155], LLMs even appear to explain their own "reasoning" and predictions.

We caution that the apparent interpretability of LLMs at the natural language level can be illusory. Recent work shows that the "reasoning" generated via chain-of-thought prompting does not always reflect the true causal mechanism responsible for the model outputs.[156] We contend that responsible use of subpopulation representative models will involve interpreting LLMs at three levels: the level of their natural language inputs and outputs, their learned representations, and their learned information-processing machinery. This is fertile ground for interdisciplinary research. Although deep learning is (at least for the moment) less commonly used in the social sciences, social scientists are nonetheless well-equipped to contribute.

[151]Tianyi Zhang et al., *BERTScore: Evaluating Text Generation with BERT*, arXiv, 2020, https://doi.org/10.48550/arXiv.1904.09675.

[152]Ethan Perez et al., *Discovering Language Model Behaviors with Model-Written Evaluations*, arXiv, 2022, https://doi.org/10.48550/arXiv.2212.09251.

[153]Alon Jacovi et al., *Stop Uploading Test Data in Plain Text: Practical Strategies for Mitigating Data Contamination by Evaluation Benchmarks*, arXiv, 2023, https://doi.org/10.48550/arXiv.2305.10160.

[154]Jason Wei et al., "Chain-of-Thought Prompting Elicits Reasoning in Large Language Models," *Advances in Neural Information Processing Systems* 35 (2022): 24824–24837.

[155]Shunyu Yao et al., *Tree of Thoughts: Deliberate Problem Solving with Large Language Models*, arXiv, 2023, https://doi.org/10.48550/arXiv.2305.10601.

We see four specific areas for interpretability research for SRMs:

1. The first is the application and evaluation of existing interpretability techniques in the SRM domain, including white-box techniques, black-box techniques, and techniques to measure and improve LLM "faithfulness" (e.g., Burns et al., 2022; Cohen et al., 2023; Tafjord, Dalvi Mishra, and P. Clark, 2022).

2. The second is developing novel interpretability methods specific to the SRM setting. Approaches based on earlier embedding models like `word2vec` have already proven constructive in recovering linguistic dimensions relating to social class[157], ideology[158], and political values[159], allowing for a novel measurement strategy of textual positions in these spaces.

3. The third area involves bridging spatial models from political science and embedding space representations. Deep networks have long been appreciated for the capacity for representation learning—the ability to learn feature representations that help to facilitate the predictive task.[160] In the SRM setting, interpreting the role of embedding spaces is assisted by well-established theoretical frameworks from political science that model political competition in spa-tial terms (e.g., Downs, 1957). More specifically, the basic space theory[161] and the holographic interpretation of ideology (e.g., Bonica, 2018) emphasize how the comprehensive, high-dimensional (action) space of political issues regularly collapses to a more practical, low-dimensional ideological (basic) space. It is in this reduced basic space where political competition takes place: where parties develop reputations, candidates jockey for position, and voters evaluate candidates. In the basic space, political debate is a struggle over language and its meaning; as Hinich and Munger (1994) write: "[T]he contest is more than one of persuasion: ultimately, the contest is decided by who gets to use their words, their conception, to describe the conflict" (p. 17). Hence, it is not surprising that researchers regularly locate one or more ideological dimensions embedded in LLMs.[162]

4. Finally, the fourth area involves taking a new angle on political psychology. In LLMs, learned representations for language, and learned machinery to process these representations operate hand-in-hand to produce the behavior of the model. Similarly, human behavior in the domain of political opinion involves not only the formation of some ideological structure, but also a navigation through that structure to produce behavior. We see an opportunity to bridge the study of representation processing in humans (i.e., political psychology) and representation processing in LLMs (i.e., "mechanistic interpretability").[163] Chain-of-thought and tree-of-thought prompting (model tuning techniques that were introduced in Section 2.3) might also be used to test models of cognitive processes underlying opinion formation and the survey response.

[156]Miles Turpin et al., *Language Models Don't Always Say What They Think: Unfaithful Explanations in Chain-of-Thought Prompting*, arXiv, 2023, https://doi.org/10.48550/arXiv.2305.04388.

[157]Austin C. Kozlowski, Matt Taddy, and James A. Evans, "The Geometry of Culture: Analyzing the Meanings of Class Through Word Embeddings," *American Sociological Review* 84, no. 5 (2019): 905–949. https://doi.org/10.1177/0003122419877135.

[158]Ludovic Rheault and Christopher Cochrane, "Word Embeddings For the Analysis Of Ideological Placement In

## 6.4. Sparse Cell Problem

One question to consider is whether LLMs will alleviate any of the fundamental challenges of opinion aggregation. One of these is the *sparse cell problem*, where adding more conditioning variables to a probability estimate (i.e., stratification) reduces the number of observations available to inform that estimate. This problem is well understood by researchers performing even a moderate amount of stratification to produce contingency tables. In Bayesian hierarchical (multilevel) models, the concept of *borrowed strength* is often used in reference to shrinkage techniques, where estimates for sparsely populated cells are "strengthened" via shrinkage to overall estimates, exchanging information between more and less densely populated cells. Mixture modeling strategies such as latent class analysis have also been proposed as methods to obtain more reliable estimates of cell estimates in sparse contingency tables.[164] Given the ubiquitous nature of the sparse cell problem and prior interest in solutions, practitioners are sure to be interested in whether LLMs offer an effective remedy.

*Multilevel regression and poststratification* (MRP) methods are currently the state-of-the-art approach to estimating subgroup opinion with survey data (see D. K. Park, Gelman, and Bafumi, 2004). These kinds of hierarchical modeling strategies represent a major advance from past approaches to dealing with sparse samples, including combining responses to similar survey items across multiple years (which requires the assumption of static attitudes) or using more easily accessible surrogate measures to proxy opinions (e.g., presidential vote). MRP uses a two-step process in which the target opinion (e.g., position on a late-term abortion ban) is modeled as a function of demographic covariates in the first stage, and those predictions extrapolated to known demographic proportions in subgroups (e.g., states or counties) in the second stage. While MRP is not a panacea for small or even moderately-sized surveys[165], machine learning has already helped to improve the MRP method by replacing regression models with more flexible supervised algorithms in the first-stage opinion prediction process (e.g., Bisbee, 2019).

LLMs sometimes produce outputs that seem highly unlikely to be the result of memorization.[166] Additionally, chat interfaces allow for powerful, flexible text generation based on ad-hoc instructions from users. The experience of working with such apparently capable systems could lead practitioners to an expectation that LLMs can overcome issues with training data sparsity via strong generalization. If this is true, LLMs would indeed offer

Parliamentary Corpora," *Political Analysis* 28, no. 1 (2020): 112–133. https://doi.org/10.1017/pan.2019.26.

[159]Emma Rodman, "A Timely Intervention: Tracking The Changing Meanings Of Political Concepts With Word Vectors," *Political Analysis* 28, no. 1 (2020): 87–111. https://doi.org/10.1017/pan.2019.23.

[160]Nicolas Le Roux and Yoshua Bengio, "Representational Power Of Restricted Boltzmann Machines And Deep Belief Networks," *Neural Computation* 20, no. 6 (2008): 1631–1649. https://doi.org/10.1162/neco.2008.04-07-510., Ian J. Goodfellow, Jonathon Shlens, and Christian Szegedy, "Explaining and Harnessing Adversarial Examples," in *3rd International Conference On Learning Representations, ICLR 2015, San Diego, CA, USA, May 7-9, 2015, Conference Track Proceedings*, eds. Yoshua Bengio and Yann LeCun (2015), https://doi.org/10.48550/arXiv.1412.6572.

[161]Melvin J. Hinich and Michael C. Munger, *Ideology and the Theory of Political Choice* (Ann Arbor: University of Michigan Press, 1994).

[162]Mohit Iyyer et al., "Political Ideology Detection Using Recursive Neural Networks," in *Proceedings of the 52nd Annual Meeting Of the Association For Computational Linguistics (Volume 1: Long Papers)* (Baltimore, MD: Association for Computational Linguistics, 2014), 1113–1122, https://doi.org/10.3115/v1/P14-1105., Ludovic Rheault and Christopher Cochrane, "Word Embeddings For the Analysis Of Ideological Placement In Parliamentary Corpora," *Political Analysis* 28, no. 1 (2020): 112–133. https://doi.org/10.1017/pan.2019.26.

[163]For a taxonomy of interpretability approaches for deep neural networks, we refer the reader to work by Räuker et al. (2023) .

[164]Drew A. Linzer, "Reliable Inference in Highly Stratified Contingency Tables: Using Latent Class Models as Density Estimators," *Political Analysis* 19, no. 2 (2011): 173–187.

[165]See Buttice and Highton (2017) for extensive simulation results regarding MRP performance.

[166]For example, explaining a sorting algorithm in the style of a 1940s gangster (https://twitter.com/goodside/status/1598129631609380864).

a solution to the sparse cell problem— sparse cell data being data that is underrepresented or not represented at all during training.

This question gets at the heart of a central debate in machine learning—to what extent can a learning system based on *correlations* in observed data claim to be a causal model? This is an important question. For correlational systems, predictive performance degrades as counterfactual inputs get further from the observed data manifold.[167] By recovering the underlying causal mechanism that produced the observed data, causal inference offers better counterfactual inference.[168] In what category do LLMs fall? Some have argued that LLMs are purely correlational systems. Recent evidence suggests that despite being trained only on text observations, on token prediction tasks, LLMs may implicitly learn to approximate some causal learning abilities. A recent study shows that GPT-4 outperforms several state-of-the-art causal inference algorithms.[169]

Empirically, LLM performance is generally not uniform across tasks, or subtasks. The figures in Puchert et al. (2023) present a clear illustration of this phenomenon. The results from Santurkar et al. (2023) show that SRM performance varies across demographics and sub-demographics. This variance of ability is often related to how much useful data is present for each task or subtask in the model's pretraining and fine-tuning data. As an example, Razeghi et al. (2022) show that the frequency of certain numbers in an LLM's pretraining data dictates the ability of the model to perform arithmetic using those numbers. Rather than learning a robust algorithm for addition that performs equally well on all numbers, LLMs perform better for numbers that were seen more frequently during training. Extrapolating from this result to the SRM setting, it is likely that SRMs will perform better for subpopulations with more representation in the pretraining data.

The "hit-or-miss" nature of LLM generalization presents a challenge for the practitioner—LLMs can be surprisingly bad on some specific tasks, or even specific task instances, despite strong overall performance in a domain. Ultimately, while LLMs do present an opportunity for counterfactual analysis, these abilities cannot be assumed for novel tasks, requiring the practitioner to validate the LLM for each task at hand.

### 6.5. Sampling Methodology

Sampling methodology is another challenging aspect of opinion aggregation. Traditional survey research has been forced to adapt to declining and differential response rates by developing methods to correct for biases in nonprobability samples. Even nonprobability samples are often prohibitively expensive for many researchers, especially when the sample must be large enough to include an adequate number of respondents from small geographic or demographic subgroups.

In the SRM setting, even establishing the representation of certain demographics in the training data is a nontrivial task. To pre-train an LLM, a sample of text documents is drawn from the internet. At the fine-tuning stage, data is either sampled from internet text (including social media posts, as in H. Jiang et al., 2022), repurposed from survey databases or other data repositories (Argyle et al. 2023), or produced in a bespoke crowdsourcing process (M. A. Bakker et al. 2022). In traditional polling, surveying, or sociological experiment design, demographics for samples are known. In the SRM setting, different sources of data offer varying amounts of demographic metadata. Despite not being *explicitly representative*, in the sense that no metadata links them to a demo-graphic of interest, some of this data will still be more or less *functionally representative*—that is, useful for inducing fidelity towards a particular demographic. Explicit demographics for a sample can be computed prior to constructing an SRM, if the requisite metadata are available. Measuring functional representativeness, or at least measuring system performance across demographics can only be done post hoc, after constructing the system, requiring practitioners to adopt the algorithmic modeling approach (see Breiman, 2001; Grimmer, M. E. Roberts, and Stewart, 2021).

[167]Gary King and Langche Zeng, "The Dangers of Extreme Counterfactuals," *Political Analysis* 14, no. 2 (2006): 131–159. https://doi.org/10.1093/pan/mpj004.

[168]Stefan Wager and Susan Athey, "Estimation and Inference of Heterogeneous Treatment Effects Using Random Forests," *Journal of the American Statistical Association* 113, no. 523 (2018): 1228–1242. https://doi.org/10.1080/01621459.2017.1319839., Susan Athey, Julie Tibshirani, and Stefan Wager, "Generalized Random Forests," *The Annals of Statistics* 47, no. 2 (2019): 1148–1178. https://doi.org/10.1214/18-AOS1709.

[169]Cheng Zhang et al., *Understanding Causality with Large Language Models: Feasibility and Opportunities*, arXiv, 2023, https://doi.org/10.48550/arXiv.2304.05524.

Representativeness is also an issue during prompting. Argyle et al. (2023) proposes to use prompting to correct the "skewed marginals" that occur during pre-training (presumably as a result of the pretraining dataset not being sampled representatively). While this shows some effectiveness, this technique is not equally effective for all groups Santurkar et al. (2023) , and may over-correct or produce response distributions with less variance than would be observed in the corresponding human subpopulation (e.g., Bisbee et al., 2023). LLMs have been shown to sometimes be sensitive to particular prompt phrasing. Hagendorff (2023) makes an analogy to convenience sampling, suggesting that experiments that use the first few prompts that come to the minds of the researchers could be made more rigorous by applying a more principled prompt selection methodology.

## 7. Conclusion

In this review, we considered how LLMs might be applied to better understand opinion at the subpopulation level. We drew together the body of literature using Large Language Models (LLMs) as Subpopulation Representative Models (SRMs)—models that approximate to some useful degree some characteristics of human subpopulations. These models have been proposed as a complement or alternative to more traditional methods for aggregating community sentiment, since they offer unique benefits including decreased cost, the potential for nuanced and open-ended analysis, and the potential for forecasting analysis to unobserved events. Just as the computational power of the "People Machine" helped the Kennedy campaign better hear the voices of neglected subgroups, the application of LLMs to opinion aggregation could be constructive and normatively desirable. However, just as Morgan warns of unintended consequences of the "People Machine", the introduction of LLMs into the fragile ecosystem between public opinion, democratic elections, and political institutions also carries first and second-order risks, many of which would be quite familiar to Thomas B. Morgan or other observers of the People Machine some half-century years ago. To make the best of this precarious transition, we encourage a precautionary, interdisciplinary effort from all relevant disciplines. We hope that practitioners will see SRMs not only as isolated technological implements, tools to acquire political advantage, or a fertile area for academic innovation. We hope that SRMs are seen as interactants in a complex system, whose introduction therein will have consequences for generations to come.

## Bibliography

AI Objectives Institute. *Introducing Talk To the City: Collective Deliberation at Scale.* 2023, https://www.talktothe.city/.

Achille, Alessandro et al. *The Information Complexity of Learning Tasks, Their Structure and Their Distance.* arXiv, 2020, https://doi.org/10.48550/arXiv.1904.03292.

Aharoni, Roee, and Yoav Goldberg. "Unsupervised Domain Clusters in Pretrained Language Models." In *Proceedings of the 58th Annual Meeting of the Association for Computational Linguistics.* Online: Association for Computational Linguistics, 2020, 7747–7763, https://doi.org/10.18653/v1/2020.acl-main.692.

Aher, Gati, Rosa I. Arriaga, and Adam Tauman Kalai. *Using Large Language Models to Simulate Multiple Humans and Replicate Human Subject*

*Studies.* arXiv, 2023, https://doi.org/10.48550/arXiv.2208.10264.

Alvarez, R. Michael, and John Brehm. *Hard Choices, Easy Answers: Values, Information, and American Public Opinion.* Princeton, NJ: Princeton University Press, 2002.

Alvarez, R. Michael, Ines Levin, and Lucas Núñez. "The Four Faces of Political Participation in Argentina: Using Latent Class Analysis to Study Political Behavior." *Journal of Politics* 79, no. 4. 2017: 1386–1402.

Alwin, Duane F., and Jon A. Krosnick. "The Reliability of Survey Attitude Measurement: The Influence of Question and Respondent Attributes." *Sociological Methods & Research* 20, no. 1. 1991: 139–181.

Amershi, Saleema et al. "Software Engineering for Machine Learning: A Case Study." In *2019 IEEE/ACM 41st International Conference On Software Engineering: Software Engineering In Practice (ICSE-SEIP).* 2019, 291–300, https://doi.org/10.1109/ICSE-SEIP.2019.00042.

Amodei, Dario et al. *Concrete Problems In AI Safety.* arXiv, 2016, https://doi.org/10.48550/arXiv.1606.06565.

Argyle, Lisa P. et al. "Out of One, Many: Using Language Models to Simulate Human Samples." *Political Analysis* 31, no. 3. 2023: 337–351.

Ashmore, Rob, Radu Calinescu, and Colin Paterson. "Assuring the Machine Learning Lifecycle: Desiderata, Methods, and Challenges." *ACM Computing Surveys* 54, no. 5. 2022: 1–39. https://doi.org/10.1145/3453444.

Athey, Susan, Julie Tibshirani, and Stefan Wager. "Generalized Random Forests." *The Annals of Statistics* 47, no. 2. 2019: 1148–1178. https://doi.org/10.1214/18-AOS1709.

Austin, Robert D. *Measuring and Managing Performance in Organizations.* New York: Dorset House, 1996.

Bahdanau, Dzmitry, Kyunghyun Cho, and Yoshua Bengio. *Neural Machine Translation by Jointly Learning to Align and Translate.* arXiv, 2016, https://doi.org/10.48550/arXiv.1409.0473.

Bai, Yuntao et al. *Training a Helpful and Harmless Assistant with Reinforcement Learning from Human Feedback.* arXiv, 2022, https://doi.org/10.48550/arXiv.2204.05862.

Bai, Yuntao et al. *Constitutional AI: Harmlessness from AI Feedback.* arXiv, 2022, https://doi.org/10.48550/arXiv.2212.08073.

Bakker, Michiel A. et al. "Fine-Tuning Language Models to Find Agreement among Humans with Diverse Preferences." *Advances in Neural Information Processing Systems* 35. 2022: 38176–38189.

Berinsky, Adam J. "Measuring Public Opinion with Surveys." *Annual Review of Political Science* 20. 2017: 309–329.

Binz, Marcel, and Eric Schulz. "Using Cognitive Psychology to Understand GPT-3." *Proceedings of the National Academy of Sciences* 120, no. 6. 2023. https://doi.org/10.1073/pnas.2218523120.

Bisbee, James. "BARP: Improving Mister P Using Bayesian Additive Regression Trees." *American Political Science Review* 113, no. 4. 2019: 1060–1065. https://doi.org/10.1017/S0003055419000480.

Bisbee, James et al. *Artificially Precise Extremism: How Internet-Trained LLMs Exaggerate Our Differences.* SocArXiv, 2023, https://doi.org/10.31235/osf.io/5ecfa.

Bommasani, Rishi et al. *On the Opportunities and Risks of Foundation Models.* arXiv, 2022, https://doi.org/10.48550/arXiv.2108.07258.

Bonica, Adam. "Inferring Roll-Call Scores from Campaign Contributions Using Supervised Machine Learning." *American Journal of Political Science* 62, no. 4. 2018: 830–848. https://doi.org/https://doi.org/10.1111/ajps.12376.

Boopathy, Akhilan et al. "Model-Agnostic Measure of Generalization Difficulty." In *Proceedings of the 40th International Conference on Machine Learning*. Eds. Andreas Krause et al. PMLR, 2023, 2857–2884, https://proceedings.mlr.press/v202/boopathy23a.html.

Bowman, Samuel R. et al. *Measuring Progress on Scalable Oversight for Large Language Models*. arXiv, 2022, https://doi.org/10.48550/arXiv.2211.03540.

Breiman, Leo. "Statistical Modeling: The Two Cultures." *Statistical Science* 16, no. 3. 2001: 199–231.

Brown, Tom et al. "Language Models Are Few-Shot Learners." *Advances in Neural Information Processing Systems* 33. 2020: 1877–1901.

Burns, Collin et al. *Discovering Latent Knowledge in Language Models Without Supervision*. arXiv, 2022, https://doi.org/10.48550/arXiv.2212.03827.

Buttice, Matthew K., and Benjamin Highton. "How Does Multilevel Regression And Poststratification Perform With Conventional National Surveys?." *Political Analysis* 21, no. 4. 2017: 449–467. https://doi.org/10.1093/pan/mpt017.

Callegaro, Mario et al. "A Critical Review of Studies Investigating the Quality of Data Obtained with Online Panels Based on Probability and Nonprobability Samples." In *Online Panel Research: A Data Quality Perspective*. Eds. Mario Callegaro et al. New York: Wiley, 2014, 23–53.

Carmines, Edward G., and James A. Stimson. *Issue Evolution: Race and the Transformation of American Politics*. Princeton, NJ: Princeton University Press, 1989.

Carmines, Edward G., and James A. Stimson. "The Two Faces of Issue Voting." *American Political Science Review* 74, no. 1. 1980: 78–91.

Caughey, Devin, and Christopher Warshaw. "Dynamic Estimation of Latent Opinion Using a Hierarchical Group-Level IRT Model." *Political Analysis* 23, no. 2. 2015: 197–211.

Cheng, Emily, Mathieu Rita, and Thierry Poibeau. *On the Correspondence between Compositionality and Imitation in Emergent Neural Communication*. arXiv, 2023, https://doi.org/10.48550/arXiv.2305.12941.

Chowdhary, Girish, Chinmay Soman, and Katherine Driggs-Campbell. *Levels of Autonomy for Field Robots*. 2020, https://www.earthsense.co/news/2020/7/24/levels-of-autonomy-for-field-robots.

Christian, Brian. *The Alignment Problem: Machine Learning and Human Values*. New York: W. W. Norton & Company, 2020.

Chu, Eric et al. *Language Models Trained on Media Diets Can Predict Public Opinion*. arXiv, 2023, https://doi.org/10.48550/arXiv.2303.16779.

Cinelli, Matteo et al. *Echo Chambers on Social Media: A Comparative Analysis*. arXiv, 2020, https://doi.org/10.48550/arXiv.2004.09603.

Cohen, Roi et al. *LM Vs LM: Detecting Factual Errors Via Cross Examination*. arXiv, 2023, https://doi.org/10.48550/arXiv.2305.13281.

Commons, Michael Lamport. "Introduction to the Model of Hierarchical Complexity." *Behavioral Development Bulletin* 13, no. 1. 2007: 1–6. https://doi.org/10.1037/h0100493.

Converse, Philip E. "The Nature of Belief Systems in Mass Publics." In *Ideology and Discontent*. Ed. David E. Apter. New York: Free Press, 1964, 206–261.

Cornesse, Carina et al. "A Review of Conceptual Approaches and Empirical Evidence on Probability and Nonprobability Sample Survey Research." *Journal of Survey Statistics and Methodology* 8, no. 1. 2020: 4–36.

Crawley, Sam, Hilde Coffé, and Ralph Chapman. "Public Opinion on Climate Change: Belief and Concern, Issue Salience and Support for Government Action." *The British Journal of Politics* *and International Relations* 22, no. 1. 2020: 102–121.

De Marchi, Scott, and Scott E. Page. "Agent-Based Models." *Annual Review of Political Science* 17. 2014: 1–20.

Dean, Walter. "Computational Complexity Theory." In *The Stanford Encyclopedia Of Philosophy*. Ed. Edward N. Zalta. Metaphysics Research Lab, Stanford University, 2021, https://plato.stanford.edu/archives/fall2021/entries/computational-complexity/.

December, John. "Units of Analysis for Internet Communication." *Journal of Communication* 46, no. 1. 1996: 14–38.

Devlin, Jacob et al. "BERT: Pre-Training of Deep Bidirectional Transformers for Language Understanding." In *Proceedings of the 2019 Conference of the North American Chapter of the Association for Computational Linguistics: Human Language Technologies*. Minneapolis, MN: Association for Computational Linguistics, 2019, 4171–4186, https://doi.org/10.18653/v1/N19-1423.

Downs, Anthony. *An Economic Theory of Democracy*. New York: Harper, Row, 1957.

El-Mhamdi, El-Mahdi et al. *SoK: On The Impossible Security Of Very Large Foundation Models*. arXiv, 2022, https://doi.org/10.48550/arXiv.2209.15259.

Feldman, Philip et al. *Polling Latent Opinions: A Method for Computational Sociolinguistics Using Transformer Language Models*. arXiv, 2022, https://doi.org/10.48550/arXiv.2204.07483.

Feldman, Philip G., Shimei Pan, and James Foulds. "The Keyword Explorer Suite: A Toolkit for Understanding Online Populations." In *Companion Proceedings of the 28th International Conference on Intelligent User Interfaces*. New York: Association for Computing Machinery, 2023, 21–24, https://doi.org/10.1145/3581754.3584122.

Feldman, Stanley, and Christopher Johnston. "Understanding the Determinants of Political Ideology: Implications of Structural Complexity." *Political Psychology* 35, no. 3. 2014: 337–358.

Feng, Shangbin et al. "From Pretraining Data To Language Models To Downstream Tasks: Tracking The Trails Of Political Biases Leading To Unfair NLP Models." In *Proceedings of the 61st Annual Meeting Of the Association For Computational Linguistics (Volume 1: Long Papers)*. Toronto, Canada: Association for Computational Linguistics, 2023, 11737–11762, https://aclanthology.org/2023.acl-long.656.

Fleishman, John A. "Types of Political Attitude Structure: Results of a Cluster Analysis." *Public Opinion Quarterly* 50, no. 3. 1986: 371–386.

Fowler, Anthony et al. "Moderates." *American Political Science Review* 117, no. 2. 2023: 643–660.

France-Presse, Agence. "Romania PM Unveils AI 'adviser' to Tell Him What People Think in Real Time." *The Guardian*. March 2023. https://www.theguardian.com/world/2023/mar/02/romania-ion-ai-government-honorary-adviser-artificial-intelligence-pm-nicolae-ciuca.

French, Robert M. "Catastrophic Forgetting in Connectionist Networks." *Trends in Cognitive Sciences* 3, no. 4. 1999: 128–135. https://doi.org/10.1016/S1364-6613(99)01294-2.

Fu, Zihao et al. *On the Effectiveness of Parameter-Efficient Fine-Tuning*. arXiv, 2022, https://doi.org/10.48550/arXiv.2211.15583.

Gall, John. *The Systems Bible: The Beginner's Guide to Systems Large and Small*. Walker, MN: General Systemantics Press, 2002.

Garimella, Kiran et al. "Political Discourse on Social Media: Echo Chambers, Gatekeepers, and the Price of Bipartisanship." In *Proceedings of the 2018 World Wide Web Conference*. Lyon, France: International World Wide Web Conferences Steering Committee, 2018, 913, https://doi.org/10.1145/3178876.3186139.

Goodfellow, Ian J., Jonathon Shlens, and Christian Szegedy. "Explaining and Harnessing Adversarial Examples." In *3rd International Conference On Learning Representations, ICLR 2015, San Diego, CA, USA, May 7-9, 2015, Conference Track Proceedings*. Eds. Yoshua Bengio and Yann LeCun. 2015, https://doi.org/10.48550/arXiv.1412.6572.

Graham, Jesse et al. "Moral Foundations Theory: The Pragmatic Validity of Moral Pluralism." In . Eds. Patricia Devine and Ashby Plant. Advances in Experimental Social Psychology. 2013, 55–130, https://doi.org/0.1016/B978-0-12-407236-7.00002-4.

Griffin, Lewis D. et al. *Susceptibility to Influence of Large Language Models*. arXiv, 2023, https://doi.org/10.48550/arXiv.2303.06074.

Grimmer, Justin, Margaret E. Roberts, and Brandon M. Stewart. "Machine Learning for Social Science: An Agnostic Approach." *Annual Review of Political Science* 24, no. 1. 2021: 1–25.

Gross, Justin H, and Daniel Manrique-Vallier. "A Mixed Membership Approach to Political Ideology." In *Handbook of Mixed Membership Models and Their Applications*. Eds. Edoardo M. Airoldi et al. Boca Raton, FL: CRC Press, 2015, 119–140.

Groves, Robert M. "Three Eras of Survey Research." *Public Opinion Quarterly* 75, no. 5. 2011: 861–871.

Gu, Yuxian et al. "PPT: Pre-Trained Prompt Tuning for Few-Shot Learning." In *Proceedings of the 60th Annual Meeting Of the Association For Computational Linguistics (Volume 1: Long Papers)*. Dublin, Ireland: Association for Computational Linguistics, 2022, 8410–8423, https://aclanthology.org/2022.acl-long.576.

Gururangan, Suchin et al. *Annotation Artifacts in Natural Language Inference Data*. arXiv, 2018, https://doi.org/10.48550/arXiv.1803.02324.

Hagendorff, Thilo. *Machine Psychology: Investigating Emergent Capabilities and Behavior in Large Language Models Using Psychological Methods*. arXiv, 2023, https://doi.org/10.48550/arXiv.2303.13988.

Hale, John. "Information-Theoretical Complexity Metrics." *Language and Linguistics Compass* 10, no. 9. 2016: 397–412. https://doi.org/10.1111/lnc3.12196.

Hastie, Trevor, Robert Tibshirani, and Jerome Friedman. *The Elements of Statistical Learning*. 2nd ed.. Springer Series in Statistics. New York: Springer, 2009.

Hendra, Richard, and Aaron Hill. "Rethinking Response Rates: New Evidence of Little Relationship between Survey Response Rates and Nonresponse Bias." *Evaluation Review* 43, no. 5. 2019: 307–330.

Herbst, Susan. *Numbered Voices: How Opinion Polling Has Shaped American Politics*. Chicago: University of Chicago Press, 1993.

Hill, Jennifer L., and Hanspeter Kriesi. "Classification by Opinion-Changing Behavior: A Mixture Model Approach." *Political Analysis* 9, no. 4. 2001: 301–324.

Hillygus, D. Sunshine. "The Practice of Survey Research: Changes and Challenges." In *New Directions in Public Opinion*. Ed. Adam J. Berinsky. New York: Routledge, 2020, 21–40.

Hinich, Melvin J., and Michael C. Munger. *Ideology and the Theory of Political Choice*. Ann Arbor: University of Michigan Press, 1994.

Howard, Philip N., and Muzammil M. Hussain. *Democracy's Fourth Wave? Digital Media and the Arab Spring*. New York: Oxford University Press, 2013.

Hu, Edward J. et al. "LoRA: Low-Rank Adaptation of Large Language Models." In *International Conference on Learning Representations*. 2022, https://openreview.net/forum?id=nZeVKeeFYf9.

Hutchinson, Ben et al. "Towards Accountability for Machine Learning Datasets: Practices

from Software Engineering and Infrastructure." In *Proceedings of the 2021 ACM Conference On Fairness, Accountability, And Transparency*. New York: Association for Computing Machinery, 2021, 560–575, https://doi.org/10.1145/3442188.3445918.

Huynh, Daniel, and Jade Hardouin. *PoisonGPT: How We Hid a Lobotomized LLM on Hugging Face to Spread Fake News*. July 9, 2023, https://blog.mithrilsecurity.io/poisongpt-how-we-hid-a-lobotomized-llm-on-hugging-face-to-spread-fake-news/.

Isler, Ozan, Onurcan Yilmaz, and Burak Dogruyol. "Activating Reflective Thinking with Decision Justification and Debiasing Training." *Judgment and Decision Making* 15, no. 6. 2020: 926–938. https://doi.org/10.1017/S1930297500008147.

Iyyer, Mohit et al. "Political Ideology Detection Using Recursive Neural Networks." In *Proceedings of the 52nd Annual Meeting Of the Association For Computational Linguistics (Volume 1: Long Papers)*. Baltimore, MD: Association for Computational Linguistics, 2014, 1113–1122, https://doi.org/10.3115/v1/P14-1105.

Jacovi, Alon et al. *Stop Uploading Test Data in Plain Text: Practical Strategies for Mitigating Data Contamination by Evaluation Benchmarks*. arXiv, 2023, https://doi.org/10.48550/arXiv.2305.10160.

Jiang, Hang et al. "CommunityLM: Probing Partisan Worldviews from Language Models." In *Proceedings of the 29th International Conference On Computational Linguistics*. Gyeongju, Republic of Korea: International Committee on Computational Linguistics, 2022, 6818–6826, https://aclanthology.org/2022.coling-1.593.

Jost, John T. et al. "Political Conservatism as Motivated Social Cognition." *Psychological Bulletin* 129, no. 3. 2003: 339–375.

Kahan, Dan M. "Ideology, Motivated Reasoning, and Cognitive Reflection." *Judgment and Decision Making* 8, no. 4. 2013: 407–424.

Kahneman, Daniel, and Shane Frederick. "Representativeness Revisited: Attribute Substitution in Intuitive Judgment." In *Heuristics and Biases: The Psychology of Intuitive Judgment*. New York: Cambridge University Press, 2002, 49–81, https://doi.org/10.1017/CBO9780511808098.004.

Kane, Michael. "Validity and Fairness." *Language Testing* 27, no. 2. 2010: 177–182. https://doi.org/10.1177/0265532209349467.

Kennedy, Courtney et al. "An Evaluation of the 2016 Election Polls in the United States." *Public Opinion Quarterly* 82, no. 1. 2018: 1–33.

Key, V.O., Jr. *Public Opinion and American Democracy*. New York: Knopf, 1961.

Kim, Junsol, and Byungkyu Lee. *AI-Augmented Surveys: Leveraging Large Language Models for Opinion Prediction in Nationally Representative Surveys*. arXiv, 2023, https://doi.org/10.48550/arXiv.2305.09620.

King, Gary, and Langche Zeng. "The Dangers of Extreme Counterfactuals." *Political Analysis* 14, no. 2. 2006: 131–159. https://doi.org/10.1093/pan/mpj004.

Kozlowski, Austin C., Matt Taddy, and James A. Evans. "The Geometry of Culture: Analyzing the Meanings of Class Through Word Embeddings." *American Sociological Review* 84, no. 5. 2019: 905–949. https://doi.org/10.1177/0003122419877135.

Krosnick, Jon A. "Government Policy and Citizen Passion: A Study of Issue Publics in Contemporary America." *Political Behavior* 12, no. 1. 1990: 59–92.

Kuang, Kai. "The Role of Interactivity in New Media-Based Health Communication: Testing the Interaction among Interactivity, Threat, and Efficacy." In *Technology and Health*. Eds.

Jihyun Kim and Hayeon Song. Academic Press, 2020, 377–397, https://doi.org/10.1016/B978-0-12-816958-2.00017-4.

Laato, Samuli et al. "AI Governance in the System Development Life Cycle: Insights on Responsible Machine Learning Engineering." In *Proceedings of the 1st International Conference On AI Engineering: Software Engineering For AI*. New York: Association for Computing Machinery, 2022, 113–123, https://doi.org/10.1145/3522664.3528598.

Ladyman, James, James Lambert, and Karoline Wiesner. "What Is a Complex System?." *European Journal for Philosophy of Science* 3. 2013: 33–67.

Lambe, Kathryn Ann et al. "Dual-Process Cognitive Interventions to Enhance Diagnostic Reasoning: A Systematic Review." *BMJ Quality & Safety* 25, no. 10. 2016: 808–820. https://doi.org/10.1136/bmjqs-2015-004417.

Le Roux, Nicolas, and Yoshua Bengio. "Representational Power Of Restricted Boltzmann Machines And Deep Belief Networks." *Neural Computation* 20, no. 6. 2008: 1631–1649. https://doi.org/10.1162/neco.2008.04-07-510.

LePore, Jill. *If Then: How the Simulmatics Corporation Invented the Future*. New York: Liveright Publishing, 2020.

Leeper, Thomas J. "Where Have the Respondents Gone? Perhaps We Ate Them All." *Public Opinion Quarterly* 83, S1. 2019: 280–288.

Lester, Brian, Rami Al-Rfou, and Noah Constant. *The Power of Scale for Parameter-Efficient Prompt Tuning*. arXiv, 2021, https://doi.org/10.48550/arXiv.2104.08691.

Lewis, Patrick et al. *Retrieval-Augmented Generation for Knowledge-Intensive NLP Tasks*. arXiv, 2021, https://doi.org/10.48550/arXiv.2005.11401.

Li, Bo et al. "Trustworthy AI: From Principles To Practices." *ACM Computing Surveys* 55, no. 9. 2023: 1–46. https://doi.org/10.1145/3555803.

Li, Ming, and Paul Vitányi. *An Introduction to Kolmogorov Compelxity and Its Applications*. 4th ed.. Texts in Computer Science. New York: Springer, 2019.

Lin, Chin-Yew. "ROUGE: A Package For Automatic Evaluation Of Summaries." In *Text Summarization Branches Out*. Barcelona, Spain: Association for Computational Linguistics, 2004, 74–81, https://aclanthology.org/W04-1013.

Lin, Zhiyu et al. *Beyond Prompts: Exploring the Design Space of Mixed-Initiative Co-Creativity Systems*. arXiv, 2023, https://doi.org/10.48550/arXiv.2305.07465.

Linzer, Drew A. "Reliable Inference in Highly Stratified Contingency Tables: Using Latent Class Models as Density Estimators." *Political Analysis* 19, no. 2. 2011: 173–187.

Liu, Pengfei et al. "Pre-Train, Prompt, and Predict: A Systematic Survey of Prompting Methods in Natural Language Processing." *ACM Computing Surveys* 55, no. 9. 2023: 1–35.

Ma, Huan et al. *Fairness-Guided Few-Shot Prompting for Large Language Models*. arXiv, 2023, https://doi.org/10.48550/arXiv.2303.13217.

Madaan, Aman et al. *Self-Refine: Iterative Refinement with Self-Feedback*. arXiv, 2023, https://doi.org/10.48550/arXiv.2303.17651.

Magee, Christopher, and Olivier de Weck. *Complex System Classification*. International Council On Systems Engineering (INCOSE), 2004, accessed July 22, 2023, https://dspace.mit.edu/handle/1721.1/6753.

Martin, Donald et al. *Extending the Machine Learning Abstraction Boundary: A Complex Systems Approach to Incorporate Societal Context*. arXiv, 2020, https://doi.org/10.48550/arXiv.2006.09663.

Mattu, Jeff et al. *Machine Bias.* 2016, accessed October 20, 2023, https://www.propublica.org/article/machine-bias-risk-assessments-in-criminal-sentencing.

McCutcheon, Allan L. "A Latent Class Analysis of Tolerance for Nonconformity in the American Public." *Public Opinion Quarterly* 49, no. 4. 1985: 474–488.

Mehrabi, Ninareh et al. "A Survey on Bias and Fairness in Machine Learning." *ACM Computing Surveys* 54, no. 6. 2021: 1–35. https://doi.org/10.1145/3457607.

Merrill, Jeremy B., and Will Oremus. "Five Points for Anger, One for a 'like': How Facebook's Formula Fostered Rage and Misinformation." *The Washington Post.* October 2021. https://www.washingtonpost.com/technology/2021/10/26/facebook-angry-emoji-algorithm/.

Messaoudi, Chaima, Zahia Guessoum, and Lotfi Ben Romdhane. "Opinion Mining in Online Social Media: A Survey." *Social Network Analysis and Mining* 12, no. 1. 2022: 25. https://doi.org/10.1007/s13278-021-00855-8.

Mikolov, Tomas et al. *Efficient Estimation of Word Representations in Vector Space.* arXiv, 2013, https://doi.org/10.48550/arXiv.1301.3781.

Morgan, Thomas B. "The People-Machine." *Harper's* 222. 1961: 53–57.

Mueller, Robert S. *Report On The Investigation Into Russian Interference In The 2016 Presidential Election.* US Department of Justice, 2019, https://media.npr.org/assets/news/2019/04/muellerreport.pdf.

Muller, Jerry Z. *The Tyranny of Metrics.* Princeton: Princeton University Press, 2018.

Munn, Luke. "Angry by Design: Toxic Communication and Technical Architectures." *Humanities and Social Sciences Communications* 7, no. 1. 2020: 1–11.

Nakajima, Yohei. *BabyAGI.* 2023, https://github.com/yoheinakajima/babyagi.

Novick, David G., and Stephen Sutton. "What Is Mixed-Initiative Interaction?." In *Proceedings of the AAAI Spring Symposium on Computational Models for Mixed Initiative Interaction.* 1997, 114–116.

Osika, Anton. *GPT-Engineer.* 2023, https://github.com/AntonOsika/gpt-engineer.

Palakodety, Shriphani, Ashiqur R. KhudaBukhsh, and J. Carbonell. *Mining Insights from Large-Scale Corpora Using Fine-Tuned Language Models.* Amsterdam: IOS Press, 2020, 1890–1897.

Papineni, Kishore et al. "BLEU: A Method for Automatic Evaluation of Machine Translation." In *Proceedings of the 40th Annual Meeting Of the Association For Computational Linguistics.* Philadelphia, PA: Association for Computational Linguistics, 2002, 311–318, https://doi.org/10.3115/1073083.1073135.

Park, David K., Andrew Gelman, and Joseph Bafumi. "Bayesian Multilevel Estimation with Poststratification: State-Level Estimates from National Polls." *Political Analysis* 12, no. 4. 2004: 375–385.

Park, Joon Sung et al. *Generative Agents: Interactive Simulacra of Human Behavior.* arXiv, 2023, https://doi.org/10.48550/arXiv.2304.03442.

Pavlik, John. *New Media Technology: Cultural and Commercial Perspectives.* 2nd ed. Boston: Pearson, 1997.

Pennington, Jeffrey, Richard Socher, and Christopher Manning. "GloVe: Global Vectors for Word Representation." In *Proceedings of the 2014 Conference on Empirical Methods in Natural Language Processing (EMNLP).* Doha, Qatar: Association for Computational Linguistics, 2014, 1532–1543, https://doi.org/10.3115/v1/D14-1162.

Perez, Ethan et al. *Discovering Language Model Behaviors with Model-Written Evaluations.* arXiv, 2022, https://doi.org/10.48550/arXiv.2212.09251.

Petropoulos, Fotios et al. "Forecasting: Theory and Practice." *International Journal of Forecasting* 38, no. 3. 2022: 705–871.

Pool, Ithiel de Sola, Robert P. Abelson, and Samuel Popkin. *Candidates, Issues, and Strategies: A Computer Simulation of the 1960 and 1964 Presidential Elections.* Cambridge, MA: MIT Press, 1964.

Puchert, Patrik et al. *LLMMaps -- A Visual Metaphor For Stratified Evaluation Of Large Language Models.* arXiv, 2023, https://doi.org/10.48550/arXiv.2304.00457.

Radford, Alec et al. *Language Models Are Unsupervised Multitask Learners.* Open AI, 2019.

Razeghi, Yasaman et al. "Impact of Pretraining Term Frequencies On Few-Shot Numerical Reasoning." In *Findings of the Association For Computational Linguistics: EMNLP 2022.* Abu Dhabi, UAE: Association for Computational Linguistics, 2022, 840–854, https://aclanthology.org/2022.findings-emnlp.59.

Rheault, Ludovic, and Christopher Cochrane. "Word Embeddings For the Analysis Of Ideological Placement In Parliamentary Corpora." *Political Analysis* 28, no. 1. 2020: 112–133. https://doi.org/10.1017/pan.2019.26.

Rodman, Emma. "A Timely Intervention: Tracking The Changing Meanings Of Political Concepts With Word Vectors." *Political Analysis* 28, no. 1. 2020: 87–111. https://doi.org/10.1017/pan.2019.23.

Rothschild, Jacob E. et al. "Pigeonholing Partisans: Stereotypes of Party Supporters and Partisan Polarization." *Political Behavior* 41, no. 2. 2019: 423–443. https://doi.org/10.1007/s11109-018-9457-5.

Räuker, Tilman et al. "Toward Transparent AI: A Survey On Interpreting The Inner Structures Of Deep Neural Networks." In *2023 IEEE Conference On Secure And Trustworthy Machine Learning (SaTML).* 2023, 464–483, https://doi.org/10.1109/SaTML54575.2023.00039.

Santurkar, Shibani et al. *Whose Opinions Do Language Models Reflect?.* arXiv, 2023, https://doi.org/10.48550/arXiv.2303.17548.

Saunders, William et al. *Self-Critiquing Models for Assisting Human Evaluators.* arXiv, 2022, https://doi.org/10.48550/arXiv.2206.05802.

Schick, Timo et al. *Toolformer: Language Models Can Teach Themselves to Use Tools.* arXiv, 2023, https://doi.org/10.48550/arXiv.2302.04761.

Seger, Elizabeth et al. *Tackling Threats to Informed Decision-Making in Democratic Societies: Promoting Epistemic Security in a Technologically-Advanced World.* London: The Alan Turing Institute, 2020.

Shavit, Yonadav. *What Does It Take to Catch a Chinchilla? Verifying Rules On Large-Scale Neural Network Training Via Compute Monitoring.* arXiv, 2023, https://doi.org/10.48550/arXiv.2303.11341.

Shelby, Renee et al. *Identifying Sociotechnical Harms of Algorithmic Systems: Scoping a Taxonomy for Harm Reduction.* arXiv, 2023, https://doi.org/10.48550/arXiv.2210.05791.

Simmons, Gabriel. "Moral Mimicry: Large Language Models Produce Moral Rationalizations Tailored to Political Identity." In *Proceedings of the 61st Annual Meeting of the Association for Computational Linguistics (Volume 4: Student Research Workshop).* Toronto, Canada: Association for Computational Linguistics, 2023, 282–297, https://doi.org/10.18653/v1/2023.acl-srw.40.

Taber, Charles S., and Lodge Milton. "Motivated Skepticism in the Evaluation of Political Beliefs." *American Journal of Political Science* 50, no. 3. 2006: 755–769.

Tafjord, Oyvind, Bhavana Dalvi Mishra, and Peter Clark. "Entailer: Answering Questions With Faithful and Truthful Chains Of Reasoning." In *Proceedings of the 2022 Conference On Empirical Methods In Natural Language Processing.* Abu Dhabi, UAE:

Association for Computational Linguistics, 2022, 2078–2093, https://aclanthology.org/2022.emnlp-main.134.

Taori, Rohan et al. *Alpaca: A Strong, Replicable Instruction-Following Model.* Stanford Center for Research on Foundation Models, 2023, https://crfm.stanford.edu/2023/03/13/alpaca.html.

Thornton, Judd R. "Ambivalent or Indifferent? Examining the Validity of an Objective Measure of Partisan Ambivalence." *Political Psychology* 32, no. 5. 2011: 863–884.

Turing, Alan M. "Computing Machinery and Intelligence." *Mind* 59. 1950: 433–460. https://doi.org/10.1093/mind/lix.236.433.

Turpin, Miles et al. *Language Models Don't Always Say What They Think: Unfaithful Explanations in Chain-of-Thought Prompting.* arXiv, 2023, https://doi.org/10.48550/arXiv.2305.04388.

Vapnik, V.N. "An Overview of Statistical Learning Theory." *IEEE Transactions on Neural Networks* 10, no. 5. 1999: 988–999. https://doi.org/10.1109/72.788640.

Vaswani, Ashish et al. *Attention Is All You Need.* arXiv, 2017, https://doi.org/10.48550/arXiv.1706.03762.

Wager, Stefan, and Susan Athey. "Estimation and Inference of Heterogeneous Treatment Effects Using Random Forests." *Journal of the American Statistical Association* 113, no. 523. 2018: 1228–1242. https://doi.org/10.1080/01621459.2017.1319839.

Waggoner, Philip D. *Modern Dimension Reduction.* Elements in Quantitative and Computational Methods for the Social Sciences. New York: Cambridge University Press, 2021, https://doi.org/10.1017/9781108981767.

Walden, David D., Garry J. Roedler, and Kevin Forsberg. "INCOSE Systems Engineering Handbook Version 4: Updating the Reference for Practitioners." *INCOSE International Symposium* 25, no. 1. 2015: 678–686.

Wallace, Eric et al. "Universal Adversarial Triggers for Attacking and Analyzing NLP." In *Proceedings of the 2019 Conference On Empirical Methods In Natural Language Processing And the 9th International Joint Conference On Natural Language Processing (EMNLP-IJCNLP).* Hong Kong: Association for Computational Linguistics, 2019, 2153–2162, https://doi.org/10.18653/v1/D19-1221.

Walther, Joseph B. et al. "Attributes of Interactive Online Health Information Systems." *Journal of Medical Internet Research* 7, no. 3. 2005. https://doi.org/10.2196/jmir.7.3.e33.

Wan, Alexander et al. *Poisoning Language Models during Instruction Tuning.* arXiv, 2023, https://doi.org/10.48550/arXiv.2305.00944.

Wang, Ben, and Aran Komatsuzaki. *GPT-J-6B: A 6 Billion Parameter Autoregressive Language Model.* 2021.

Watanabe, Sumio. "Singular Learning Theory." In *Algebraic Geometry and Statistical Learning Theory.* Ed. Sumio Watanabe. Cambridge Monographs on Applied and Computational Mathematics. New York: Cambridge University Press, 2009, 158–216, https://doi.org/10.1017/CBO9780511800474.007.

Wei, Jason et al. *Finetuned Language Models Are Zero-Shot Learners.* arXiv, 2022, https://doi.org/10.48550/arXiv.2109.01652.

Wei, Jason et al. "Chain-of-Thought Prompting Elicits Reasoning in Large Language Models." *Advances in Neural Information Processing Systems* 35. 2022: 24824–24837.

Weidinger, Laura et al. "Taxonomy of Risks Posed by Language Models." In *2022 ACM Conference On Fairness, Accountability, And Transparency.* New York: Association for Computing Machinery, 2022, 214–229, https://doi.org/10.1145/3531146.3533088.

Whittaker, Joe et al. "Recommender Systems and the Amplification of Extremist Content." *Inter-*

net Policy Review 10, no. 2. 2021: 1–29. https://doi.org/10.14763/2021.2.1565.

Yao, Shunyu et al. *Tree of Thoughts: Deliberate Problem Solving with Large Language Models.* arXiv, 2023, https://doi.org/10.48550/arXiv.2305.10601.

Yao, Shunyu et al. *ReAct: Synergizing Reasoning and Acting in Language Models.* arXiv, 2023, https://doi.org/10.48550/arXiv.2210.03629.

Zaller, John. *The Nature and Origins of Mass Opinion.* New York: Cambridge University Press, 1992.

Zaller, John, and Stanley Feldman. "A Simple Theory of the Survey Response: Answering Questions Versus Revealing Preferences." *American Journal of Political Science* 36, no. 3. 1992: 579–616.

Zhang, Cheng et al. *Understanding Causality with Large Language Models: Feasibility and Opportunities.* arXiv, 2023, https://doi.org/10.48550/arXiv.2304.05524.

Zhang, Tianyi et al. *BERTScore: Evaluating Text Generation with BERT.* arXiv, 2020, https://doi.org/10.48550/arXiv.1904.09675.

Zhang, Zihan et al. "Is Neural Topic Modelling Better Than Clustering? An Empirical Study on Clustering with Contextual Embeddings for Topics." In *Proceedings of the 2022 Conference of the North American Chapter of the Association for Computational Linguistics: Human Language Technologies.* Seattle, WA: Association for Computational Linguistics, 2022, 3886–3893, https://doi.org/10.18653/v1/2022.naacl-main.285.

Zhong, Ruiqi et al. *Goal Driven Discovery of Distributional Differences Via Language Descriptions.* arXiv, 2023, https://doi.org/10.48550/arXiv.2302.14233.

janus. *Simulators.* LessWrong, 2022, accessed October 19, 2023, https://www.lesswrong.com/posts/vJFdjigzmcXMhNTsx/simulators.

van der Palm, Daniël W., L. Andries van der Ark, and Jeroen K. Vermunt. "Divisive Latent Class Modeling as a Density Estimation Method for Categorical Data." *Journal of Classification* 33, no. 1. 2016: 52–72.

*ION - Primul Consilier Cu Inteligență Artificială Al Guvernului.* 2023, https://ion.gov.ro/.